\documentclass[11pt,a4paper]{article}
\pdfoutput=1

\usepackage{jheppub}
\usepackage[rgb]{xcolor}
\usepackage{color}
\usepackage{amsmath,amsfonts,amssymb,mathrsfs,graphicx,bbm,dsfont,booktabs,mathtools,braket,lmodern,tikz,slashed}

\DeclareMathAlphabet{\mathpzc}{OT1}{pzc}{m}{it}

\newcommand{\overbar}[1]{\mkern 1.5mu\overline{\mkern-1.5mu#1\mkern-1.5mu}\mkern 1.5mu}

\newcommand{\bea}{\begin{eqnarray}}
\newcommand{\eea}{\end{eqnarray}}
\def\be{\begin{equation}}
\def\ee{\end{equation}}
\newcommand{\bei}{\begin{itemize}}
\newcommand{\eei}{\end{itemize}}
\newcommand{\bee}{\begin{enumerate}}
\newcommand{\eee}{\end{enumerate}}

\newcommand{\alg}[1]{\mathfrak{#1}}

\newcommand{\psu}{\alg{psu}}

\def\ads{{\rm AdS}_5\times {\rm S}^5}

\def\ads{{\rm AdS}_5\times {\rm S}^5}

\def\am{{\rm am}}
\def\am0{{\rm am}_0}

\DeclareMathAlphabet{\mathsc}{T1}{lmr}{m}{scsl}

\expandafter\def\expandafter\bfseries\expandafter{\bfseries\ifmmode\else\boldmath\fi}
\expandafter\def\expandafter\mdseries\expandafter{\mdseries\ifmmode\else\unboldmath\fi}
\expandafter\def\expandafter\normalfont\expandafter{\normalfont\ifmmode\else\unboldmath\fi}

\definecolor{grey}{rgb}{0.4,0.4,0.5}
\definecolor{darkgreen}{rgb}{0,0.5,0}
\definecolor{darkred}{rgb}{0.6,0.0,0}
\definecolor{lightbrown}{rgb}{1,0.9,0.8}
\definecolor{brown}{rgb}{0.6,0.3,0.3}
\definecolor{darkblue}{rgb}{0,0,0.8}
\definecolor{darkmagenta}{rgb}{0.5,0,0.5}

\title{Yang-Baxter deformations, AdS/CFT, and twist-noncommutative gauge theory}

\author{Stijn J. van Tongeren}

\affiliation{Institut f\"ur Mathematik und Institut f\"ur Physik, Humboldt-Universit\"at zu Berlin, IRIS Geb\"aude, Zum Grossen Windkanal 6, 12489 Berlin, Germany}

\emailAdd{svantongeren@physik.hu-berlin.de}

\abstract{We give an AdS/CFT interpretation to homogeneous Yang-Baxter deformations of the $\ads$ superstring as noncommutative deformations of the dual gauge theory, going well beyond the canonical noncommutative case. These homogeneous Yang-Baxter deformations can be of so-called abelian or jordanian type. While abelian deformations have a clear interpretation in string theory and many already had well understood gauge theory duals, jordanian deformations appear novel on both counts. We discuss the symmetry structure of the deformed string from the uniformizing perspective of Drinfeld twists and indicate that this structure can be realized on the gauge theory side by considering theories on various noncommutative spaces. We then conjecture that these are the gauge theory duals of our strings, modulo subtleties involving singularities. We support this conjecture by a brane construction for two jordanian examples, corresponding to noncommutative spaces with $[x^-\stackrel{\star}{,}x^i] \sim x^i$ ($i=1,2$). We also discuss $\kappa$-Minkowski type deformations of $\ads$, one of which may be the gravity dual of gauge theory on spacelike $\kappa$-Minkowski space.}

\begin{document}

\begin{flushright}\small{HU-EP-15/27\\HU-MATH-15/08}\end{flushright}

\maketitle

\section{Introduction}

Integrability has led to important insights in many areas of physics, including the AdS/CFT correspondence \cite{Maldacena:1997re}. Since integrability provides us with powerful tools to study the superstring on $\ads$ and its dual planar $\mathcal{N}=4$ supersymmetric Yang-Mills theory (SYM) \cite{Arutyunov:2009ga,Beisert:2010jr}, there has been considerable interest in understanding to what extent they can be applied beyond this maximally supersymmetric case. One example of such an extension is planar $\beta$ deformed SYM and the associated Lunin-Maldacena background \cite{Lunin:2005jy,Frolov:2005ty,Frolov:2005dj}. More recently it was understood that the string sigma model can be deformed in a variety of ways while manifestly preserving integrability, and in the present paper we would like to address the AdS/CFT interpretation of large classes of them.

We will be considering deformations of the $\ads$ string based on the construction of \cite{Delduc:2013qra}.\footnote{This generalizes the earlier results of \cite{Klimcik:2002zj,Klimcik:2008eq,Delduc:2013fga}.} In its original form this gives a quantum deformation of the $\ads$ model \cite{Arutyunov:2013ega,Delduc:2014kha} whose possible interpretation in terms of string theory and AdS/CFT remains elusive.\footnote{In the $\mathrm{AdS}_2\times \mathrm{S}^2$ and $\mathrm{AdS}_3\times \mathrm{S}^3$ cases some progress has been made on the supergravity front \cite{Lunin:2014tsa}, and a suitable maximal deformation limit gives the $\ads$ mirror model \cite{Arutynov:2014ota,Arutyunov:2014cra,Arutyunov:2014jfa}, which is a solution of supergravity. In general, the maximal deformation limit is closely related to $\mathrm{dS}_5 \times \mathrm{H}^5$ \cite{Delduc:2013qra,Delduc:2014kha,Hoare:2014pna,Arutyunov:2014cra,Arutyunov:2014jfa}. Moreover, interesting links to the $\lambda$ deformation of the non-abelian T dual of the $\ads$ string \cite{Sfetsos:2013wia,Hollowood:2014qma,Demulder:2015lva}, which have supergravity embeddings \cite{Sfetsos:2014cea}, have recently been uncovered \cite{Vicedo:2015pna,Hoare:2015gda}. These $\lambda$ deformations generalize the earlier work of \cite{Balog:1993es}.}
This ``Yang-Baxter'' deformation is based on a solution of the modified (inhomogeneous) classical Yang-Baxter equation, but the construction of \cite{Delduc:2013qra} can be extended to integrable deformations based on the homogeneous classical Yang-Baxter equation \cite{Kawaguchi:2014qwa}. In this setting the homogeneous classical Yang-Baxter equation (CYBE) has many solutions, giving an abundance of integrable deformations.

All currently known solutions of the CYBE fall into two classes, known as abelian and jordanian respectively. The abelian solutions mostly have nice interpretations in terms of string theory and AdS/CFT, including for example the Lunin-Maldacena background mentioned above \cite{Matsumoto:2014nra}, and the gravity dual of canonical noncommutative SYM \cite{Matsumoto:2014gwa}, which is hence integrable. In fact, it is heuristically clear that abelian solutions correspond to TsT transformations (Melvin twists) \cite{vanTongeren:2015soa}, establishing their status in string theory.\footnote{TsT stands for T duality - shift - T duality, where for us a TsT transformation $(x,y)_\beta$ means we T dualize in $x$, shift $y$ by $\beta$ times the T dual field, and T dualize back.} Jordanian deformations are more mysterious, and necessarily deform anti-de Sitter space since the required algebraic structure is not compatible with the sphere ($\mathfrak{su}(4)$).\footnote{Upon including fermions the story may be more involved, possibly allowing a new class of deformations.} It is not obvious that the result of these deformations is always a string background, though the metric and B field of the only thus far investigated case are part of one \cite{Kawaguchi:2014fca,Matsumoto:2014ubv}. Furthermore, it is not known whether jordanian deformations have an AdS/CFT interpretation, and if so what it is. Given the nice interpretation of many of their abelian cousins, we would like to shed light on this, and attempt to give these theories an interpretation in terms of the AdS/CFT (gauge/gravity) correspondence.

In this paper we will firstly argue that all abelian and jordanian deformations of the $\ads$ string result in Drinfeld twists \cite{DrinfeldTwistRef} of the symmetry algebra. This provides a unified picture of various known abelian deformations which extends to the new jordanian ones. We then use this structure to conjecture an AdS/CFT interpretation for all such twisted models, by considering deformations of $\mathcal{N}=4$ SYM that similarly carry such twisted Hopf algebras. Since the symmetries of $\mathrm{AdS}_5$ correspond to spacetime symmetries of $\mathcal{N}=4$ SYM, these deformations lie in the realm of noncommutative field theory,\footnote{Noncommutative field theory has a rich history and relevance which we will not attempt to cover here. We refer the reader to the extensive reviews \cite{Douglas:2001ba,Szabo:2001kg}.} in the twist formalism \cite{Chaichian:2004za,Chaichian:2004yh}, see also e.g. \cite{Aschieri:2005zs,Aschieri:2006ye,Aschieri:2007sq} and the reviews \cite{Szabo:2006wx,Dimitrijevic:2014dxa}. In this formalism we start from a standard Hopf algebra based on vector fields on spacetime, a representation of which is carried by the algebra of fields (functions on spacetime). This algebra of fields is then deformed by a Drinfeld twist, so that it naturally carries the representation of a twisted Hopf algebra, resulting in a noncommutative space. While likely technically involved, we believe it should in principle be possible to construct (supersymmetric) gauge theories on such noncommutative spaces using the methods developed in \cite{Seiberg:1999vs,Madore:2000en,Jurco:2001rq}, see also \cite{Douglas:2001ba,Szabo:2001kg,Aschieri:2007sq,Dimitrijevic:2014dxa} and references therein, and the discussion below. Based on matching symmetry structures we conjecture that twisted deformations of $\ads$ of jordanian as well as abelian type, generically represent gravity duals to this type of noncommutative gauge theories. This picture is supported by a number of established abelian cases, unified in a framework that extends to jordanian deformations.

Noncommutative field theories are known to arise in the low energy physics of open strings stretching between D branes \cite{Connes:1997cr,Douglas:1997fm,Chu:1998qz,Schomerus:1999ug,Seiberg:1999vs}, and ideally instances of our conjecture should be supported by pictures of this type. In the spirit of \cite{Maldacena:1997re} we would need brane geometries which in a suitable low energy limit can on the one hand be described in terms of open strings that give us our noncommutative field theories, and on the other as closed strings in a near horizon geometry matching our deformations of $\ads$. Such pictures are already established in various abelian cases, and we will provide two examples in the new class of jordanian theories. However, already in the abelian case the situation is subtle when dealing with canonical noncommutativity in time, where the appropriate decoupling limit actually gives a noncritical open string theory instead of a field theory \cite{Seiberg:2000ms,Gopakumar:2000na}. This indicates that there is a special set of cases in which our general symmetry based conjecture may break down. In fact, while we can construct the abelian deformation of $\ads$ that we would naively associate to this noncommutativity in time, it is singular. Based on this and other examples, we believe that singularities may signal subtleties or the breakdown of our general conjecture, which we otherwise expect to hold.

As jordanian examples with a brane picture, we consider two formally similar but physically distinct jordanian deformations of $\ads$ which we can give an explicit embedding in supergravity. We then give deformations of the D3 brane metric that on the one hand have a low energy description in terms of gravity on these jordanian deformed $\ads$ spaces, and on the other hand show that we are dealing with D3 branes in a plane wave geometry with a nonconstant but exact B field. The type of noncommutativity associated to this background follows from the general results of \cite{Cornalba:2001sm}, see also \cite{Chu:2002in} and the earlier work \cite{Alekseev:1999bs,Alekseev:2000fd,Ho:2000fv}. This matches the type of noncommutativity predicted by our Drinfeld twist, which we find to be of the kind
\begin{equation*}
[x^-\stackrel{\star}{,}x^j] = i a \, x^j, \hspace{10pt} j=1,2,
\end{equation*}
where $x^-$ is one of the light cone coordinates in the $(x^0,x^3)$ plane. The second deformation has a minus sign for $x^2$. The first model has sixteen real supercharges - the maximal number since we break manifest conformal symmetry - while the second has none.

With our notion of possible dual CFTs in mind, we can try to reverse the question and investigate other noncommutative structures. Namely, if the noncommutative structure of a potential field theory we are interested can be represented via a twist, we can use the corresponding $r$ matrix to generate a deformation of $\ads$ as a candidate gravitational dual.

As an example, one noncommutative space that has been studied extensively over the last decades is generalized $\kappa$-Minkowski space \cite{Majid:1994cy} with $[x^\mu\stackrel{\star}{,}x^\nu]=i\kappa^{-1}(a^\mu x^\nu - a^\nu x^\mu)$. Timelike $a^\mu$ gives true $\kappa$-Minkowski space. Generalized $\kappa$-Minkowski space carries the action of the $\kappa$-Poincar\'e group \cite{Lukierski:1991pn,Lukierski:1992dt}, with structures that appear in what is known as doubly special relativity \cite{AmelinoCamelia:2000mn,AmelinoCamelia:2000ge} as discussed in \cite{KowalskiGlikman:2004qa}. There has been considerable interest in constructing field theories on $\kappa$-Minkowski space, see e.g. \cite{AmelinoCamelia:2001fd,Dimitrijevic:2003wv,Agostini:2004cu,Schenkel:2010sc,Meljanac:2011cs} and \cite{Dimitrijevic:2003pn,Dimitrijevic:2005xw} for gauge theories in particular. The $\kappa$-Poincar\'e group is a proper quantum group however - except in the null case, see e.g. \cite{Borowiec:2013lca,Juric:2015hda} - which is not a structure we expect to reproduce here. As such, in the present setting we can consider only the null case of the $\kappa$-Poincar\'e $r$ matrices. However, generic $\kappa$-Minkowski space also arises in a different, twisted setting, which might actually be more amenable to field theory constructions \cite{Dimitrijevic:2011jg,Dimitrijevic:2014dxa}. This is not possible within the Poincar\'e algebra \cite{Borowiec:2013gca}, but fortunately it is with the Poincar\'e-Weyl algebra \cite{Borowiec:2008uj,Dimitrijevic:2014dxa}, hence the conformal algebra.

We use the corresponding $r$ matrices and the null $\kappa$-Poincar\'e one to find deformations of $\ads$. Since $\kappa$-Minkowski noncommutativity involves time however, we should perhaps expect difficulties in giving these models an AdS/CFT interpretation. In line with this, similarly to the naive canonical temporal-spatial noncommutative case mentioned earlier, the ``timelike $\kappa$-Minkowski deformation of $\ads$'' is singular, as are the two null cases which include the admissible $\kappa$-Poincar\'e case. Interestingly, by contrast the spacelike deformation is regular, leading us to conjecture that it may be the gravity dual of SYM on spacelike $\kappa$-Minkowski space in the twisted sense, though we cannot claim direct support. We have not attempted to embed the metrics and B fields of these $\kappa$-Minkowski deformations in supergravity - with the exception of one null case which is closely related to our main examples - or consequently tried to investigate a possible D brane picture, leaving this for future investigation. This and other gravitational dual descriptions may be able to give new insights into twist-noncommutative field theories in general.

This paper is organized as follows. In the next section we briefly recall the deformed string sigma model and the interpretation of the deformation as a nonlocal gauge transformation. In section \ref{sec:twistedsymmetry} we recall general aspects of Drinfeld twists and their realizations. Our results start from section \ref{sec:twistinAdSCFT}, where we give a Drinfeld twist interpretation to homogeneous Yang-Baxter deformations of the string, and use this structure to formulate our conjecture regarding their field theory duals, including subtleties and their relation to regularity of the deformed spacetimes. We then provide a brane construction illustrating our general conjecture with two jordanian examples in section \ref{sec:braneconstruction}. In section \ref{sec:kappaminkowskiandsingularities} we take a look at deformations related to $\kappa$-Minkowski space. We conclude with further discussion and associated open questions.

While this paper was in preparation we learned that reference \cite{Matsumoto:2015ypa} considered some of the $r$ matrices we consider in section \ref{sec:kappaminkowskiandsingularities}, in the context of sigma models on four dimensional Minkowski space.

\section{Homogeneous Yang-Baxter deformations of the $\ads$ string}
\label{sec:YangBaxterString}

Let us begin by recalling the definition of the deformed string sigma models we will be working with, and the interpretation of the deformation as a nonlocal gauge transformation.

\subsection*{The deformed action}

The deformations of the $\ads$ superstring that we consider are described by the action \cite{Delduc:2013qra,Kawaguchi:2014qwa} \footnote{Here $T$ is the effective string tension related to the 't Hooft coupling $\lambda$ as $T = \sqrt{\lambda}/{2\pi}$, $h$ is the world sheet metric, $\epsilon^{\tau\sigma}=1$, $A_\alpha = g^{-1} \partial_\alpha g$ with $g\in \mathrm{PSU}(2,2|4)$, $\mathrm{sTr}$ denotes the supertrace, and $d_\pm = \pm P_1 + 2 P_2 \mp P_3$ where the $P_i$ are the projectors onto the $i$th $\mathbb{Z}_4$ graded components of the semi-symmetric space $\mathrm{PSU}(2,2|4)/(\mathrm{SO}(4,1)\times \mathrm{SO}(5))$ (super $\ads$).}
\begin{equation}
\label{eq:defaction}
S = -\tfrac{T}{2} \int d\tau d\sigma \tfrac{1}{2}(\sqrt{h} h^{\alpha \beta} -\epsilon^{\alpha \beta}) \mathrm{sTr} (A_\alpha d_+ J_\beta)
\end{equation}
where $J=(1-\eta R_g \circ d_+)^{-1}(A)$ with $R_g(X)=g^{-1} R(g Xg^{-1}) g$. Setting $\eta=0$ ($R=0$) gives the undeformed $\ads$ superstring action of \cite{Metsaev:1998it}, which is famously integrable \cite{Bena:2003wd}. Now, provided $R$ is antisymmetric,
\begin{equation}
\mathrm{sTr}(R(m) n) = -\mathrm{sTr}(m R(n)),
\end{equation}
and satisfies the classical Yang-Baxter equation (CYBE)
\begin{equation}
\label{eq:CYBE}
[R(m),R(n)] - R([R(m),n] + [m,R(n)])=0,
\end{equation}
these deformed models are classically integrable as well, owing to the on shell flatness of the deformed current $J$. They also have $\kappa$ symmetry. The operator $R$ is a linear map from a given Lie (super)algebra $\mathfrak{g}$ to itself, which can be conveniently represented as
\begin{equation}
R(m)= (r)^{ij} t_i\, \mathrm{sTr}(t_j m) = \mathrm{sTr}_2(r (1\otimes m))
\end{equation}
for some anti-symmetric matrix $r$
\begin{equation}
r=(r)^{ij} t_i \wedge t_j = \frac{1}{2}(r)^{ij}  (t_i \otimes t_j -t_j \otimes t_i)
\end{equation}
with sum implied, and the $t_i$ are the generators of $\mathfrak{g}$. In our case $\mathfrak{g} = \mathfrak{psu}(2,2|4)$ \cite{Delduc:2013qra,vanTongeren:2015soa}. We will refer to both the operator $R$ and its matrix representation $r$ as the $r$ matrix, where the latter satisfies the CYBE in the form
\begin{equation}
[r_{12},r_{13}]+[r_{12},r_{23}]+[r_{13},r_{23}] =0.
\end{equation}
Here $r_{mn}$ denotes the matrix realization of $r$ acting in spaces $m$ and $n$ in a tensor product, not to be confused with the matrix elements $(r)^{ij}$ with respect to a basis of $\mathfrak{g}$. This is an admittedly abstract construction, but given an $r$ matrix we can expand the above action to get a sigma model on an explicit background. We briefly indicate the general procedure, including our algebra conventions and group parametrization, in appendix \ref{app:algebraetc}. To directly provide context, the $r$ matrix \cite{Matsumoto:2014gwa}
\begin{equation}
\label{eq:rcanNC}
r = a^2 p_2 \wedge p_3,
\end{equation}
where the $p_i$ denote the translation generators of the conformal group, produces the string sigma model defined on the gravity dual of canonical noncommutative SYM \cite{Hashimoto:1999ut,Maldacena:1999mh}, where $a$ is the parameter used in \cite{Maldacena:1999mh}. Moreover, the $r$ matrix \cite{Matsumoto:2014nra}
\begin{equation}
\label{eq:Rmatrixbetadef}
r = -\epsilon^{ijk} \tfrac{\hat{\gamma}_i}{8} h_j \wedge h_k,
\end{equation}
sums running from one to three, where the $h_k$ denote the Cartans of $\mathfrak{su}(4)$, gives the three parameter generalization \cite{Frolov:2005dj} of the Lunin-Maldacena background \cite{Lunin:2005jy} ($\hat{\gamma}_i=\hat{\beta}$).\footnote{This and similar statements throughout this paper have strictly speaking only been verified at the bosonic level. While there is no conceptual difference for the fermions, evaluating the deformed fermionic action is technically involved.}

\subsection*{The deformation as a nonlocal gauge transformation}
\label{subsec:nonlocalgaugetf}

Any model of the above type is related to the undeformed $\ads$ superstring by a nonlocal gauge transformation, as originally found for the Lunin-Maldacena background \cite{Frolov:2005dj,Alday:2005ww}. Specifically, there is a nonlocal gauge transformation that relates the deformed current $J$ to the undeformed current $A$ \cite{Matsumoto:2014cja}, see also \cite{Vicedo:2015pna}. Working in conformal gauge, we introduce worldsheet light cone coordinates $\sigma^\pm = (\tau \pm \sigma)/2$ and light cone components of the deformed current $J$ as\footnote{While we follow \cite{Matsumoto:2014cja}, not all details are identical.}
\begin{equation}
J_\pm \equiv \frac{1}{1\mp \eta R_g \circ d_\pm} A_\pm.
\end{equation}
These can be expressed as
\begin{equation}
\label{eq:Jfromgtilde}
J_\pm = \tilde{g}^{-1} \partial_\pm \tilde{g},
\end{equation}
where $\tilde{g}$ is related to $g$ as
\begin{equation}
\tilde{g} = F^{-1} g,
\end{equation}
by the twist $F$
\begin{equation}
\label{eq:twistfunctiondef}
F(\tau, \sigma) = \mbox{Pexp}\left(-\int_0^\sigma d\hat{\sigma} J_\sigma^g\right) Z,
\end{equation}
with $J_\sigma^g = g J_\sigma g^{-1} -\partial_\sigma g g^{-1}$. In other words, we have
\begin{equation}
\label{eq:currentgaugetf}
J_\pm = \mathscr{G} A_\pm \mathscr{G}^{-1} -\partial_\pm \mathscr{G} \mathscr{G}^{-1},
\end{equation}
where $\mathscr{G}=g^{-1} F g$ is the nonlocal gauge transformation. There is still some freedom left in the constant matrix $Z$, which as we will soon see, mixes the conserved charges of the model. For future reference we note that
\begin{equation}
\label{eq:JgviaR}
J_\pm^g  = g(J_\pm - A_\pm)g^{-1} =  g \left(\frac{\pm \eta R_g \circ d_\pm}{1\mp \eta R_g \circ d_\pm} A_\pm\right) g^{-1} = \pm \eta R(g d_\pm J_\pm g^{-1}).
\end{equation}

\subsection*{Conserved charges}

The equations of motion of the model take the form \cite{Matsumoto:2014cja}
\begin{equation}
\partial_\alpha \Lambda^\alpha +[J_\alpha,\Lambda^\alpha]=0,
\end{equation}
where we have introduced $d_\pm J_\pm = 2 \Lambda_\pm$, i.e.
\begin{equation}
\label{eq:lambdadef}
\Lambda^\alpha = \sqrt{h} h^{\alpha\beta} J_\beta^{(2)} - \tfrac{1}{2}\epsilon^{\alpha\beta}(J^{(1)}_\beta - J^{(3)}_\beta).
\end{equation}
These expressions are analogous to the ones for the undeformed model \cite{Arutyunov:2009ga}, just with $A(g)$ replaced by $J(\tilde{g})$. Given relation \eqref{eq:Jfromgtilde}, we can define the conserved current
\begin{equation}
\label{eq:conservedcurrent}
k^\alpha \equiv \tilde{g} \Lambda^a \tilde{g}^{-1},
\end{equation}
which transforms adjointly under changes of $Z$. Because we are working in a nonlocal and nonperiodic setting, we cannot generate conserved charges for our closed string from this conserved current, except in the undeformed limit. Still, we could consider our sigma model on a line with boundary conditions that $g$ becomes constant at large $|\sigma|$, so that we could construct conserved charges out of $k$ as
\begin{equation}
\label{eq:conscharges}
Q \equiv \int_{-\infty}^{\infty} k^\tau.
\end{equation}
These charges should generate $\mathfrak{psu}(2,2|4)$, which is however (partially) broken by the nonperiodic boundary conditions for the actual string that we consider.  In order to discuss the hidden symmetry algebras of these deformed models, we need to recall some relevant algebraic structures.

\section{Twisted symmetry}

\label{sec:twistedsymmetry}

Deformed symmetry algebras play an important role in integrable models, and (both) are intimately tied to the theory of quantum groups. The standard quantum deformation of a Lie (bi)algebra is associated to the Drinfeld-Jimbo solution of the \emph{modified} classical Yang-Baxter equation, with quantum affine algebras as the corresponding quantization of classical affine algebras. In line with this, the original inhomogeneous Yang-Baxter deformation of the $\ads$ superstring of \cite{Delduc:2013qra} - based on the Drinfeld-Jimbo $r$ matrix - results in a (standard) quantum deformation of $\mathfrak{psu}(2,2|4)$ \cite{Delduc:2014kha}. Now, solutions of the classical Yang-Baxter equation give rise to nonstandard quantizations instead, which can be represented as Drinfeld twists \cite{DrinfeldTwistRef}. Since we are considering models based on such solutions, by analogy we expect to be dealing with a twisted $\mathfrak{psu}(2,2|4)$ Hopf algebra, and implicitly its associated twisted Yangian. In section \ref{sec:twistinAdSCFT} we will make this more explicit and use this structure to conjecture associated field theory duals, but first we would like to recall the relevant algebraic structures, and their possible realization in sigma models and four dimensional field theory.

\subsection{Hopf algebras and Drinfeld twists}

The standard way to associate a Hopf algebra to a Lie algebra $\mathfrak{g}$ is to take the universal enveloping algebra $\mathcal{U}(\mathfrak{g})$ and endow it with a coproduct
\begin{equation}
\begin{aligned}
\Delta(X) &= X \otimes 1 + 1 \otimes X, \mbox{ for } X \in \mathfrak{g},\\
\Delta(1) &= 1\otimes1.
\end{aligned}
\end{equation}
This coproduct is manifestly bilinear, as well as coassociative
\begin{equation}
(\Delta \otimes 1)\Delta = (1\otimes \Delta) \Delta.
\end{equation}
We then define a co-unit $\epsilon:\mathfrak{g} \rightarrow \mathbb{C}$ as
\begin{equation}
\epsilon(1)=1, \hspace{20pt} \epsilon(X)=0,
\end{equation}
so that we have a coalgebra. By construction the comultiplication
\begin{equation}
\begin{aligned}
\Delta(XY) &=\Delta(X) \Delta(Y),\\
\epsilon(XY) &=\epsilon(X) \epsilon(Y),\\
\end{aligned}
\end{equation}
is an algebra homomorphism of $\mathcal{U}(\mathfrak{g})$ ($\Delta([X,Y])=[\Delta(X),\Delta(Y)]$), so we are dealing with a bialgebra. Finally, to turn this into a Hopf algebra we define an antipode map $s$
\begin{equation}
s(X) = -X.
\end{equation}
For details and generalizations we refer to the book \cite{Chari}.

\paragraph{Drinfeld twists.} A Drinfeld twist $\mathcal{F}$ is now an invertible element of $\mathcal{U}(\mathfrak{g}) \otimes \mathcal{U}(\mathfrak{g})$ which satisfies the cocycle condition \cite{DrinfeldTwistRef,Reshetikhin:1990ep}
\begin{equation}
(\mathcal{F} \otimes 1)(\Delta \otimes 1) \mathcal{F} = (1 \otimes \mathcal{F})(1\otimes \Delta) \mathcal{F},
\end{equation}
and the normalization condition
\begin{equation}
(\epsilon \otimes 1) \mathcal{F} =(1\otimes \epsilon) \mathcal{F} = 1 \otimes 1.
\end{equation}
Since $\mathcal{F}$ should represent a deformation, we also want
\begin{equation}
\label{genFexpansion}
\mathcal{F} = 1\otimes 1 +  \alpha \mathcal{F}^{(1)} + \mathcal{O}(\alpha^2),
\end{equation}
where $\alpha$ is a deformation parameter. Let us now express $\mathcal{F}$ as a sum of terms in $\mathcal{U}(\mathfrak{g})\otimes \mathcal{U}(\mathfrak{g})$
\begin{equation}
\mathcal{F} = f^\beta \otimes f_\beta, \hspace{10pt} \mathcal{F}^{-1} = \bar{f}^\beta \otimes \bar{f}_\beta,
\end{equation}
where $f^\beta$, $f_\beta$, $\bar{f}^\beta$, and $\bar{f}_\beta$ all denote in principle distinct elements of $\mathcal{U}(\mathfrak{g})$, and we have an implicit (infinite) sum over $\beta$. We can then modify the original coproduct and antipode $s$ of our Hopf algebra to
\begin{equation}
\label{eq:twistedhopfalgebra}
\begin{aligned}
\Delta_\mathcal{F} (X) & = \mathcal{F} \Delta(X) \mathcal{F}^{-1},\\
s_\mathcal{F}(X) & = f^\alpha s(f_\alpha) s(X) s(\bar{f}^\beta) \bar{f}_\beta.
\end{aligned}
\end{equation}
The cocycle condition guarantees that the twist preserves coassociativity of the coproduct which will come back later.

\paragraph{Drinfeld twists and $r$ matrices.} Drinfeld twists correspond one to one to classical $r$ matrices in the following sense \cite{DrinfeldTwistRef,Giaquinto:1994jx}. Firstly, the classical $r$ matrix constructed as
\begin{equation}
r_{12} = \tfrac{1}{2}(\mathcal{F}^{(1)}_{12} - \mathcal{F}^{(1)}_{21}),
\end{equation}
solves the CYBE. Secondly, any twists that have the same classical $r$ matrix result in equivalent quantizations (deformations) of the algebra. Thirdly, a twist exists for any solution of the CYBE, though an explicit construction is not known in general. We will refer to twists with $\mathcal{F}^{(1)}_{12}=r_{12}$ as \emph{$r$-symmetric}. In a general integrable model, a twist changes the quantum $R$ matrix and monodromy matrix as
\begin{equation}
\label{eq:rmatrelation}
R_{12}\rightarrow R^\mathcal{F}_{12} = \mathcal{F}_{21} R_{12} \mathcal{F}_{12}^{-1}.
\end{equation}
To the author's knowledge there are two known basic types of Drinfeld twists.

\paragraph{Abelian twists} are associated to abelian $r$ matrices, which are of the form
\begin{equation}
r = \alpha_{ij} a^i \wedge a^j,
\end{equation}
where all $a^k \in \mathfrak{g}$ mutually commute. In our conventions, the associated Drinfeld-Reshetikhin twist \cite{Reshetikhin:1990ep} is given by
\begin{equation}
\label{eq:abeliantwist}
\mathcal{F} = e^{-i r},
\end{equation}
which manifestly satisfies the cocycle and normalization conditions.

\paragraph{Jordanian twists} are associated to the Borel subalgebra of $\mathfrak{g}$, where we take two generators $h$ and $e$ with $[h,e]=e$ to form\footnote{Note that this structure is not compatible with a compact Lie algebra like $\mathfrak{su}(4)$. Of course we could have equally well chosen $h$ and $f$ with $[h,f]=-f$.}
\begin{equation}
r = \beta h\wedge e.
\end{equation}
In a matrix realization where $e^2 =0$, we would have $r^3 =0$.\footnote{It is possible to have abelian $r$ matrices with a matrix realization where $r^3=0$, these are sometimes referred to as abelian-jordanian $r$ matrices \cite{Kawaguchi:2014qwa}.} A compact expression for a representative of the associated twist is
\begin{equation}
\label{eq:jordFnonsymm}
\mathcal{F} = e^{h \otimes y}, \hspace{20pt} y = \log (1 - i\beta e).
\end{equation}
Note that this twist is not $r$-symmetric, but an equivalent $r$-symmetric version exists \cite{DrinfeldTwistRef}, with $\mathcal{F}^{-1}$ given by \cite{Giaquinto:1994jx} (see also \cite{Tolstoy:2008zz})
\begin{equation}
\label{eq:Fsymm}
\mathcal{F}^{-1} = \sum_{m=0}^\infty \frac{1}{m!} \left(\frac{-i \beta}{2}\right)^m \sum_{s=0}^m (-1)^s \binom{m}{s} e^{m-s} h^{\langle s\rangle} \otimes e^s h^{\langle m-s\rangle},
\end{equation}
where
\begin{equation}
h^{\langle k\rangle} = h(h+1)\ldots (h+k-1), \hspace{10pt} k \in \mathbb{N}^+
\end{equation}
and $h^{\langle0\rangle}=1$. These twists satisfy the cocycle and normalization conditions.\footnote{In fact, both jordanian and abelian twists satisfy simpler versions of the cocycle condition, see e.g. \cite{Kulish:1998be}. When verifying the cocycle condition in a matrix realization with $e^2 =0$ we should keep in mind that $\Delta(e^2) = \Delta(e) \Delta (e)$.} In contrast to the abelian case, flipping the sign of the deformation parameter does not manifestly invert the twist. We should additionally note that there are extended jordanian twists \cite{Kulish:1998be}, see also \cite{Tolstoy:2008zz}, built out of the above abelian and jordanian building blocks. Some examples we will consider below fall into this class, but let us not get lost in further details. Regarding both these remarks, we emphasize that due to the above-mentioned theorems by Drinfeld, the essence of the deformation is captured by the $r$ matrix. Finally, note that $h$ cannot carry a physical (length) scale, only $e$ can, while in the abelian case both generators can.

\subsection{Realizations}

$\mathrm{PSU}(2,2,|4)$ represents target space (super)symmetries of the $\ads$ string - global internal symmetries - and the associated generators are realized in the sigma model via conserved charges. A hypothetical Drinfeld twist of the string should twist the Hopf algebra built on these conserved charges, and as we will demonstrate in section \ref{subsec:twistfunctionfixing} at the level of the monodromy matrix, this is precisely what happens when deforming the action with an $r$ matrix that solves the CYBE.

For $\mathcal{N}=4$ SYM on the other hand, $\mathrm{PSU}(2,2,|4)$ represents spacetime and R symmetry, and its generators are realized on the algebra of fields (functions) defined on (super)spacetime via vector fields. It is perhaps not immediately clear how twisted symmetry is to be realized here, but this can be determined from the underlying Hopf algebraic structure. The algebra of functions carries the vector field representation of the standard Hopf algebra built on $\mathfrak{psu}(2,2|4)$, and it is possible to induce a deformation of this algebra from a deformation of the Hopf algebra. This leads us into the realm of noncommutative geometry, which under the Moyal-Weyl correspondence is precisely described via a deformation of the algebra of functions on a manifold. In our case it will be Drinfeld twists that induce a deformation the algebra of (smooth) functions on Minkowski space.\footnote{Our examples of deformations of the $\ads$ string will be naturally defined in the Poincar\'e patch, which is why we focus on Minkowski space. Similar concepts should apply to global anti-de Sitter space and $\mathbb{R}\times \mathrm{S}^3$, though the fact that manifest conformal invariance is broken makes this point somewhat subtle, as mentioned in e.g. \cite{Beisert:2005if}.} Let us describe this in a bit more detail, see e.g. \cite{Szabo:2006wx} for a review.

\paragraph{Drinfeld twists and star products.} The space of functions on a manifold forms a module for the Lie algebra of vector fields on this manifold. We can give the universal enveloping algebra of vector fields on Minkowski space, $\mathcal{U}(T\mathcal{M})$, the structure of a Hopf algebra by taking
\begin{equation}
\begin{aligned}
\Delta(\xi) &= \xi \otimes 1 + 1 \otimes \xi, &
\Delta(1) &= 1 \otimes 1,\\
\epsilon(\xi) &= 0, &
s(\xi) & = -\xi,
\end{aligned}
\end{equation}
where $\xi = \xi^\mu \partial_\mu$, just like we did above for a finite dimensional Lie algebra $\mathfrak{g}$. We say that this Hopf algebra naturally acts on the algebra of functions because its algebra and coalgebra structure are respected. At the algebraic level, multiplication in $\mathcal{U}(T\mathcal{M})$ is compatible with the action of vector fields on functions
\begin{equation}
(\xi \zeta)(f) = \xi(\zeta(f)),
\end{equation}
while at the coalgebra level we simply have the product rule
\begin{equation}
\label{eq:prodrule}
\xi(fg)= \xi(\mu(f\otimes g)) = \mu(\Delta(\xi)(f \otimes g)) = \xi(f)g +f\xi(g)
\end{equation}
where $\mu(a\otimes b)=ab$ is just the usual product of functions.

We can now Drinfeld twist this Hopf algebra as in eqs. \eqref{eq:twistedhopfalgebra}, i.e.
\begin{equation}
\begin{aligned}
\Delta_\mathcal{F} (\xi) = \mathcal{F} \Delta(\xi) \mathcal{F}^{-1}, \,\,\,\, \,\,\,\,s_\mathcal{F}(\xi) = f^\alpha s(f_\alpha) s(\xi) s(\bar{f}^\beta) \bar{f}_\beta.
\end{aligned}
\end{equation}
The algebra of vector fields remains unchanged, as does the counit, but the coproduct and antipode have changed. This Hopf algebra no longer acts on functions on $\mathcal{M}$, and we can ask what type of deformation of the algebra of functions carries a representation of this Hopf algebra. Not surprisingly, this turns out to be a twist-deformed function algebra: we take functions on $\mathcal{M}$ but with the twisted product
\begin{equation}
\mu_\mathcal{F}(f\otimes g) = \mu \circ \mathcal{F}^{-1} (f \otimes g),
\end{equation}
so that as in eqn. \eqref{eq:prodrule} above we have a ``product rule''
\begin{equation}
\xi(\mu_\mathcal{F}(f\otimes g)) = \mu_\mathcal{F} \Delta_{\mathcal{F}}(\xi) (f\otimes g).
\end{equation}
We will denote this twisted product by a Groenewold-Moyal star product
\begin{equation}
f\star g \equiv \mu \circ \mathcal{F}^{-1} (f \otimes g),
\end{equation}
where the cocycle condition now ensures associativity. We will encounter examples of twisted products below. The noncommutative structure of spacetime can now be read off from
\begin{equation}
\label{eq:starcommdef}
[x\stackrel{\star}{,}y] \equiv x \star y - y \star x.
\end{equation}
Note that if we want to be able to get a Hermitian star product in the sense
\begin{equation}
\overbar{f \star g} = \bar{g} \star \bar{f}
\end{equation}
we need to work with the $r$-symmetric version of a jordanian twist. Field theories defined on this type of noncommutative space would carry a twisted Hopf algebra structure, matching the one our deformed strings realize, as we will now show.

\section{Drinfeld twisting AdS/CFT}
\label{sec:twistinAdSCFT}

Now that we have recalled the necessary background material, in this section we will argue that all homogeneous Yang-Baxter deformations result in models with Drinfeld twisted symmetries. Then, based on the possibility of realizing similar twisted symmetries in terms of noncommutative deformations of SYM, we will conjecture that the resulting theories are AdS/CFT (gauge/gravity) dual, giving an AdS/CFT interpretation to generic homogeneous Yang-Baxter deformations of the $\ads$ superstring.

\subsection{The twist function and gauge fixing}
\label{subsec:twistfunctionfixing}

Each deformation of our string arises from a particular $r$ matrix solving the CYBE. As indicated in the previous section, such $r$ matrices are in one-to-one correspondence with Drinfeld twists. Furthermore, since we are dealing with integrable deformations of the $\ads$ superstring and its Yangian symmetry, we should find some deformation (quantization) of this symmetry. Since this deformation is fixed by its leading order structure, we will only concern ourselves with this order.\footnote{\label{footnote:long}It is not obvious to the author that we can always fix $Z$ (cf. eqn \eqref{eq:twistfunctiondef}) so that the twist function manifestly gives an ($r$-symmetric) Drinfeld twist. Investigating this would require extensive studies of the classical dynamics of our deformed models, not to even mention the quantum case. Moreover, the all order gauge fixing of $Z$ depends on the $r$ matrix, while to date there is no classification of allowed $r$ matrices. Nevertheless, the structure is unambiguously fixed at leading order. This structure has been worked out to all orders in particular cases, namely for the Lunin-Maldacena deformation as an abelian example \cite{Frolov:2005dj,Alday:2005ww}, see also e.g. \cite{vanTongeren:2013gva} for the quantum story, and beyond string theory for the Schr\"odinger deformation of the $\mathrm{AdS}_3$ sigma model as a jordanian example \cite{Kawaguchi:2013lba}, see also \cite{Matsumoto:2015jja}.}

To explicitly demonstrate and identify the twist relevant for a given model, we will consider its monodromy matrix - the generator of its conserved charges. This monodromy matrix is given by
\begin{equation}
M = \mbox{Pexp}\left(-\int_{-\pi}^{\pi} d\hat{\sigma} L^g_\sigma\right)
\end{equation}
where $L^g_\sigma$ is the spatial component of the Lax connection of these models, which we take built on $J^g$ instead of $J$ but is otherwise given in \cite{Kawaguchi:2014qwa}. What will be important for us is that by the gauge transformation discussed in section \ref{sec:YangBaxterString}, this monodromy matrix is gauge equivalent to its undeformed counterpart $M_0$,\footnote{At the bosonic level this readily follows by combining the discussion of section \ref{subsec:nonlocalgaugetf} with the approach of \cite{Alday:2005ww}. Eqn. \eqref{eq:currentgaugetf} tells us that the Lax connection based on $J$ transforms as a gauge field under $\mathscr{G}$. Working based on $J^g$ instead - $dg g^{-1}$ as opposed to $g^{-1} dg$ in the undeformed setting - effectively strips $g$ and $g^{-1}$ off of $\mathscr{G}$, leaving $F$. The fermionic analysis is more involved, but we have no doubt it should go through. In any case in this paper we are only directly concerned with the bosonic part of the deformed models.}
\begin{equation}
\label{eq:monodromyrelation}
M = F(\pi)^{-1} M_0 F(-\pi) = Z^{-1} \mbox{Pexp}\left(\int_{-\pi}^\pi d\hat{\sigma} J_\sigma^g \right) M_0 Z,
\end{equation}
showing how the deformation acts on (twists) the original symmetry algebra. We will now interpret this deformation in the spirit of quantum groups.

Comparing eqs. \eqref{eq:monodromyrelation} and \eqref{eq:rmatrelation} it is natural to relate $F(\pi)$ to $\mathcal{F}_{21}^{-1}$, where we note that the quantum spaces $1$ and $2$ in the sigma model correspond to the (matrix) algebra $\mathfrak{psu}(2,2|4)$ ($\mathfrak{su}(2,2|4)$) and (classical) field space respectively. By now we of course expect to find exactly the classical $r$ matrix used to deform our model. To see this, let us suggestively rewrite $J_\sigma^g$ as
\begin{equation}
J_\sigma^g = \tfrac{1}{2}(J_+^g - J_-^g) = \eta R(g (\Lambda_+ + \Lambda_-) g^{-1}) = - 2\eta R( g \tilde{g}^{-1} k^\tau \tilde{g} g^{-1} ),
\end{equation}
cf. eqs. \eqref{eq:JgviaR}, \eqref{eq:lambdadef}, and \eqref{eq:conservedcurrent}. We can then evaluate the twist function \eqref{eq:twistfunctiondef} to leading order to find
\begin{equation}
\mathcal{F}(\pi) =  \mbox{Pexp}\left(-\int_{-\pi}^\pi d\hat{\sigma} J_\sigma^g\right) Z = \left( 1 + 2\eta R(Q_0) + \mathcal{O}(\eta^2)\right) Z.
\end{equation}
Here $Q_0$ are the conserved charges of the undeformed model cf. eqn. \eqref{eq:conscharges} (now on $[-\pi,\pi)$), generating $\mathfrak{psu}(2,2|4)$. To complete the comparison we fix $Z$ as
\begin{equation}
Z = 1 - \eta R(Q_0) + \mathcal{O}(\eta^2),
\end{equation}
which follows by noting that for an $r$-symmetric quantum twist
\begin{equation}
\mathcal{F}_{12} = 1 + \alpha r_{12} +\mathcal{O}(\alpha^2) =  \mathcal{F}_{21}^{-1} \sim F(\pi),
\end{equation}
since $r_{12}=-r_{21}$. In this gauge for $Z$, the leading order expansion of the twist function is the $r$ matrix: the supertrace in $R$ picks out individual charges that act on classical fields taking values in $\mathfrak{psu}(2,2|4)$ via the Poisson bracket. At the quantum level we therefore effectively need to multiply the $r$ matrix by $-i$, which amounts to replacing the deformation parameter $\eta$ by $-i \eta$ in the twist. Hence at the quantum level we expect to have
\begin{equation}
\mathcal{F}= 1 - i \eta r + \mathcal{O}(\eta^2).
\end{equation}
In short, we have shown that at leading order the twist function represents a Drinfeld twist, which identifies the relevant algebraic structure of our integrable model to all orders. From here on out we incorporate the deformation parameter(s) in the $r$ matrix, and therefore effectively set $\eta=1$.

\subsection{Twisted gauge theory and AdS/CFT}
\label{subsec:mainconjecture}

Our deformed strings correspond to Drinfeld twists of the original $\ads$ superstring, and we have seen above that it is possible to deform (the spacetime underlying) $\mathcal{N}=4$ SYM so as to realize the same Drinfeld twisted symmetry algebras.\footnote{We presume it is possible to construct these field theories, at least to say leading order in the deformation parameter(s).} As string theory in $\ads$ is dual to SYM, and we can deform both theories in formally the same way, we conjecture that our deformed strings (generically) represent gravity duals to noncommutative versions of SYM obtained by the corresponding Drinfeld twist. This provides a unified perspective on various known deformations of AdS/CFT in the abelian case, and importantly gives an AdS/CFT interpretation to jordanian deformations of the string.

\paragraph{Abelian deformations.} Many instances of this general conjecture were already known in the abelian case. This is not surprising, given the interpretation of abelian deformations as TsT transformations in string theory \cite{vanTongeren:2015soa}.  To start with, the deformation of the superpotential of SYM that turns it into $\beta$ deformed SYM \cite{Leigh:1995ep} can be represented by means of a star product in $\mathrm{SU}(4)$ field space \cite{Lunin:2005jy}. The $r$ matrix \eqref{eq:Rmatrixbetadef} gives the corresponding quantum twist, upon representing the Cartan generators of $\mathrm{SU}(4)$ via the R charges. Closer to the present context, the quantum twist associated to the $r$ matrix \eqref{eq:rcanNC} for canonical noncommutative SYM indeed results indeed in
\begin{equation}
\label{eq:standardNC}
[x^2\stackrel{\star}{,}x^3] \sim i a^2,
\end{equation}
upon realizing $\mathfrak{su}(2,2)$ through vector fields on $\mathbb{R}^{1,3}$, cf. appendix \ref{app:su22vec}. Similar structures arise in other theories obtained by TsT transformations, such as dipole theories ($r \sim h_i \wedge p_j$ \cite{Matsumoto:2015uja}), see e.g. \cite{Dasgupta:2001zu}, and the noncommutativity obtained by twisting in polar coordinates ($r \sim m_{12}\wedge p_3$) of e.g. \cite{Hashimoto:2005hy}. Our picture also confirms the ``guess'' for the star product of abelian Cartan-based deformations of global $\mathrm{AdS}_5$ given in \cite{Beisert:2005if}.

\paragraph{Subtleties and singularities.} Our conjecture is motivated by underlying symmetry structures, and further supported by the various concrete abelian instances just mentioned. However, there are exceptions to the rule. Namely, while the canonical spatially noncommutative case of eqn. \eqref{eq:standardNC} is ok, the usual argument that gives a noncommutative field theory description to open strings in a low energy limit breaks down in the temporal-spatial case. This is due to a critical value of the electric field ($B_{0i}$) on the brane \cite{Seiberg:2000ms,Gopakumar:2000na}. Beyond this value string pair production destabilizes the background, while we would need to cross it to get a finite noncommutative $\alpha^\prime\rightarrow0$ limit. This critical value is reflected by a pole in the open string noncommutativity parameter $\theta$. To determine whether similar critical values exist in other deformed theories, we would need to know the corresponding deformation of the D3 brane action, which we do not know in general. Still, we can readily construct the deformation of $\ads$ that we would naively associate to $[x^0\stackrel{\star}{,}x^1]\sim a^2$. The resulting geometry is just a formal TsT transformation $(x^1,x^0)_{a^2}$ of $\ads$ (in flat space this generates the desired electric field), which due to the shift of time by a spatial coordinate generates a naked singularity at $z \sim a$. This is not the gravity dual of the noncritical string theory \cite{Gopakumar:2000na}. We believe this singularity may be an indication that our considerations break down, or may be the counterpart to fundamental problems with a naively constructed dual field theory, though we should note that it is apparently possible to formulate unitary field theories with noncommutativity involving time \cite{Doplicher:1994tu,Bahns:2002vm}. In either case the singularity might indicate that the dual field theory interpretation breaks down. In line with this, canonical null-spatial noncommutativity is fine and corresponds to a regular deformation of $\ads$ \cite{Aharony:2000gz}.

\paragraph{Noncommutativity involving time.} Even the direct appearance of time in a noncommutative structure is not necessarily a problem, as shown by the model of Hashimoto and Sethi \cite{Hashimoto:2002nr}. Their noncommutative field theory and associated gravity dual are obtained by TsT transforming a stack of D3 branes, and doing a (singular) coordinate change, hence we expect this model to fit in our framework. In \cite{Matsumoto:2015ypa} it was noted that the four dimensional metric that the branes see can be reproduced by their truncated Yang-Baxter deformation of Minkowski space. Working in terms of target space light cone coordinates $x^\pm = (x^0\pm x^3)/\sqrt{2}$, we have checked that the corresponding abelian $r$ matrix
\begin{equation}
r= a m_{1-} \wedge p_2,
\end{equation}
reproduces also the corresponding gravity dual as a deformation of $\ads$, upon identifying $x^+ = a^{-1} y^+$, $x^1 = y^+ \tilde{y}$, $x^- = a y^- + \tfrac{a}{2} y^+ \tilde{y}^2$, $x^2=-\tilde{z}$ \cite{Hashimoto:2002nr,Matsumoto:2015ypa}, and $a$ as the $\tilde{R}$ of \cite{Hashimoto:2002nr}. The associated noncommutative field theory has
\begin{equation}
\label{eq:HSNC}
[x^1\stackrel{\star}{,}x^2] = i a x^+, \hspace{20pt} [x^-\stackrel{\star}{,}x^2] = i a x^1,
\end{equation}
which our Drinfeld twist picture beautifully postdicts. While this structure unequivocally involves time by containing both light cone coordinates, this theory does arise from an appropriate low energy open string limit \cite{Hashimoto:2002nr}. The associated gravity dual is regular.

\paragraph{General deformations.}  Based on the above discussion, there does not appear to be an obvious pattern that determines whether a given noncommutative structure is admissible in the present context. Based on the exceptional abelian case of a constant electric field, we expect there may be subtleties with our conjecture when a given deformation generates a singularity in the background, but otherwise we expect it to hold. In particular, jordanian deformations of $\ads$ should generically be dual to noncommutative deformations of SYM. The associated noncommutative structures need not look more involved than e.g. the one in eqn. \eqref{eq:HSNC}. As this dual picture for jordanian deformations is entirely new, we would like to support it in the spirit of \cite{Maldacena:1997re}. Let us do this in two examples.

\section{Jordanian twists and branes}
\label{sec:braneconstruction}

The first model we will consider is the jordanian deformation of $\ads$ first studied in \cite{Kawaguchi:2014fca,Matsumoto:2014ubv}. This deformation arises from the $\mathfrak{su}(2,2|4)$ $r$ matrix \cite{vanTongeren:2015soa}
\begin{equation}
\label{eq:mainRmatrixexample}
r = a(D - m_{+-}) \wedge p_-,
\end{equation}
where the $p$, $m$, and $D$, denote momenta, Lorentz transformations, and the dilatation generator respectively, cf. appendix \ref{app:su22vec}, with $[D - m_{+-},p_-]=2 p_-$. As a deformation of Poincar\'e AdS, the resulting metric and B field are given by
\begin{equation}
\label{eq:mainJordanianDefAdS5}
\begin{aligned}
ds^2 = \, &   \frac{-2 dx^+ dx^- + dx_i dx^i + dz^2}{z^2}-\frac{a^2(z^2+x_i x^i) (dx^+)^2}{z^6} + d\Omega_5^2\\
B = \, & -\frac{a(x^1 dx^1 + x^2 dx^2 - z dz )}{z^4} \wedge dx^+,
\end{aligned}
\end{equation}
where the sums run over $i=1,2$, and we note that the $z dz$ term in the B field is a total derivative.  Reducing to $\mathrm{AdS}_3$ by dropping $x_1$ and $x_2$, this gives the Schr\"odinger deformation of $\mathrm{AdS}_3$ mentioned in footnote \ref{footnote:long}. A supergravity solution containing the metric and B field of eqs. \eqref{eq:mainJordanianDefAdS5} was found in \cite{Kawaguchi:2014fca}, and later shown to match the result of two TsT transformations combined with an S duality \cite{Matsumoto:2014ubv}.\footnote{The two TsT transformations produce a $C_2$ Ramond-Ramond potential from the $C_4$ potential of $\ads$, which upon S duality becomes the needed B field. It seems unlikely that we can obtain this B field directly from a (sequence) of TsT transformations, since it is nonzero in a non-isometric direction ($\sqrt{x_i x^i}$).}

Now up to an overall sign of the B field (except on the total derivative), the same bosonic model arises from the extended $r$ matrix
\begin{equation}
r = a\left((D - m_{+-}) \wedge p_- + 2 m_{-i}\wedge p^i\right),
\end{equation}
which is known as the $\kappa$-Weyl $r$ matrix \cite{Lukierski:2002ii}, where $a$ is typically denoted $\kappa^{-1}$. Importantly, the fermions do distinguish these $r$ matrices: the model associated to eqn. \eqref{eq:mainRmatrixexample} has 16 real supercharges, while the $\kappa$-Weyl one has none.\footnote{The manifest symmetry algebra of a given deformed model is spanned by the $t \in \mathfrak{psu}(2,2|4)$ for which $R([t,x])=[t,R(x)]$ for all $x \in \mathfrak{psu}(2,2|4)$ \cite{vanTongeren:2015soa}.} We can actually drop either of the two additional terms in the $r$ matrix and still get a solution of the CYBE, though adding them with arbitrary parameters does not work. Let us focus on
\begin{equation}
\label{eq:mainRmatrixexample2}
r^\prime = a\left((D - m_{+-}) \wedge p_- + 2 m_{-2}\wedge p_2 \right),
\end{equation}
which will give our second model. These additional terms only flip signs in the B field, i.e.
\begin{equation}
\label{eq:Bprime}
B^\prime =  -a \frac{x^1 dx^1 - x^2 dx^2 - z dz}{z^4}\wedge dx^+,
\end{equation}
but leave the metric invariant. This deformation preserves no supersymmetry, and breaks rotational symmetry in the $(x^1,x^2)$ plane compared to the model above.

In contrast to the first model, it is harder to imagine reproducing this geometry by TsT transformations and S dualities; we need a sum of squares of $x^1$ and $x^2$ in the metric, but a difference of squares in the B field. Still, given the relatively simple form of the supergravity equations of motion in this case, we found a modified solution that incorporates this model, also at the D3 brane level. Hence, we will be able to analyze this second model just like the first, while it has a more generic structure. Importantly, we cannot be sure that these solutions of supergravity correspond to the deformed coset models without reading off all background fields from the worldsheet action. That being said, the essential parts of our results rely only on the metric and B field, and are insensitive to this point provided an appropriate supergravity solution exists.

Let us now find the corresponding deformations of the D3 brane background.

\subsection{Branes and low energy limits}

Since the isometries involved in the TsT transformations used to go from $\ads$ to the space of eqn. \eqref{eq:mainJordanianDefAdS5} are isometries of the full D3 brane background, we can directly apply them there. To get to \eqref{eq:mainJordanianDefAdS5} from $\ads$ we should do two TsT transformations $(x^-,\zeta)_{b}$ and $(x^-,\chi)_{\pm b}$ followed by an S duality transformation, where $\zeta$ is the polar angle in the $(x^1,x^2)$ plane and $\chi$ is the $\mathrm{S}^1$ coordinate of $\mathrm{S}^5$ viewed as an $\mathrm{S}^1$ fibration over $\mathbb{CP}^2$ \cite{Matsumoto:2014ubv}.\footnote{The sign choice on $(x^-,\chi)_{\pm b}$ only affects $C_2$ (the fermions), we present the $+$ case.} Applying these transformations to the D3 brane background (see e.g. \cite{Maldacena:1999mh}) gives
\begin{align}
ds^2 & = \frac{1}{\sqrt{f}}(-2dx^+ dx^- + dx_i dx^i - b^2(r^2 + f^{-1}x_i x^i) (dx^+)^2) + \sqrt{f}(dr^2 + r^2 d\Omega_5^2), \nonumber \\
B & = -b f^{-1}(x^1 dx^1 + x^2 dx^2)\wedge dx^+, \hspace{8pt} C_2  = b \left(f^{-1}(x^2 dx^1 - x^1 dx^2) -r^2(d\chi + \omega)\right)\wedge dx^+,\nonumber\\
C_4 & = C_4^{0}, \label{eq:defD3brane}
\end{align}
where $C_4^0$ is the undeformed potential, sums run over $i=1,2$,
\begin{equation}
f= 1 + \frac{{\alpha^\prime}^2 R^4}{r^4},
\end{equation}
and we have reinstated units in the conventions of \cite{Maldacena:1999mh}. Our coordinates on $\mathrm{S}^5$ and in particular $\omega$ are defined in appendix \ref{app:fibration}. The dilaton is constant. Note that $b$ has units of inverse length. This solution is symmetric under $x^\pm,b \rightarrow -x^\pm,-b$. The deformed D3 brane solution relevant for the $r^\prime$ deformation is obtained by replacing $B$ and $C_2$ by\footnote{This solution can be generalized to complement the version of eqn. \eqref{eq:defD3brane} obtained by taking the TsT parameters different. Intuitive explanations for the factor of $\sqrt{3}$ are welcome.}
\begin{equation}
\begin{aligned}
B  & = -b f^{-1}(x^1 dx^1 - x^2 dx^2)\wedge dx^+, \\
C_2  & = -b \left(f^{-1}(x^2 dx^1 + x^1 dx^2) - \tfrac{2}{\sqrt{3}}r^2(d\chi + \omega)\right)\wedge dx^+.
\end{aligned}
\end{equation}

\paragraph{Open strings.} Asymptotically far away we can see the original geometry the branes were placed in; given the factor $r^2 (dx^+)^2$, we rescale $x^- \rightarrow w^2 x^-, x^i\rightarrow w x^i, r \rightarrow w r$, and consider the limit $w \rightarrow \infty$. Up to an overall scale $w^2$, in Cartesian coordinates we find
\begin{equation}
\begin{aligned}
ds^2 &=-2dx^+ dx^- + dx_l dx^l - b^2 (x_l x^l) (dx^+)^2,\\
C_2 &=  b \sum_{k=1}^4 (x^{2k-1} dx^{2k} - x^{2k} dx^{2k-1})\wedge dx^+,\\
B &= -b (x^1 dx^1\pm x^2 dx^2) \wedge dx^+,
\end{aligned}
\end{equation}
where the sum in $l$ now runs over eight $x$s. This is nothing but a plane wave supported by a RR three form, with a B field that is exact but not constant.\footnote{The B field cannot be gauged away in directions along a brane, in the sense that doing so would introduce a gauge field on the brane with equivalent effects.} The plane wave potential breaks translational symmetry in particular for the center of mass of the branes. We can now look at the effective geometry seen by the open strings stretching between branes placed in this background. The metric $G$ and noncommutativity parameter $\theta$ are obtained from the closed string metric $g$ and B field\footnote{Note that our conventions differ by a factor of $2\pi \alpha^\prime (i)$ on the B field from those of \cite{Seiberg:1999vs}.} via \cite{Seiberg:1999vs,Cornalba:2001sm}
\begin{equation}
\begin{aligned}
G & = (g +  B)^{-1} g (g - B)^{-1},\\
\theta & = -2 \pi \alpha^\prime (g + B)^{-1} B (g - B)^{-1}.
\end{aligned}
\end{equation}
giving
\begin{equation}
\begin{aligned}
G_{mn} & = \eta_{mn} - \delta_{m+}\delta_{n+} b^2 \sum_{k=3}^8 x_k x^k,\\
\theta & = -2 \pi \alpha^\prime b (x^1 dx^1 \pm x^2 dx^2) \wedge dx^-,
\end{aligned}
\end{equation}
To get a finite result in the $\alpha^\prime \rightarrow 0$ limit \cite{Seiberg:1999vs} we rescale $x^\pm \rightarrow {\alpha^\prime}^{\pm1} x^\pm$ to find
\begin{equation}
\theta = -2 \pi b (x^1 dx^1 \pm x^2 dx^2) \wedge dx^-.
\end{equation}
In other words, the low energy theory of open strings stretching between D3 branes in this geometry should correspond to a noncommutative version of $\mathcal{N}=4$ SYM, with noncommutativity of the type
\begin{equation}
\label{eq:mainNCexampleYM}
[x^-\stackrel{\star}{,}x^j] = 2 \pi i b x^j,
\end{equation}
for $j=1,2$, with a minus sign for $x^2$ in the $r^\prime$ case.

\paragraph{Closed strings.} If now instead we consider the near horizon low energy limit by replacing $r\rightarrow \alpha^\prime R^2/z$ and $b \rightarrow a/\alpha^\prime$ and taking $\alpha^\prime \rightarrow 0$, by construction we get our deformed $\ads$ metric of eqn. \eqref{eq:mainJordanianDefAdS5}
\begin{equation}
(\alpha^\prime R^2)^{-1} ds^2 =    \frac{-2 dx^+ dx^- + dx_i dx^i + dz^2}{z^2}- \frac{a^2 R^4 (z^2+x_i x^i) (dx^+)^2}{z^6} + d\Omega_5^2
\end{equation}
with the B field of eqs. \eqref{eq:mainJordanianDefAdS5} and \eqref{eq:Bprime}
\begin{equation}
(\alpha^\prime R^2)^{-1} B  =  a R^2 \frac{(x^1 dx^1 \pm x^2 dx^2)}{z^4} \wedge dx^+,
\end{equation}
In our Drinfeld twist picture for the first model we get\footnote{Here we are working with $\mathcal{F}^{-1} = 1 + i r + \mathcal{O}(a^2)$ in the $r$-symmetrized form \eqref{eq:Fsymm}. Higher order terms do not contribute to this commutator given the realization of the $p_\mu$ as differential operators.}
\begin{equation}
[x^\mu\stackrel{\star}{,}x^\nu] = \mu_\mathcal{F}(x^\mu \otimes x^\nu-x^\nu \otimes x^\mu) = 2 i a R^2 \mu ((D -m_{+-}) \wedge p_-(x^\mu \otimes x^\nu)),
\end{equation}
which gives
\begin{equation}
\label{eq:mainNCexampleSTRING}
[x^-\stackrel{\star}{,}x^i] = i a R^2 x^i,
\end{equation}
cf. appendix \ref{app:su22vec}. This nicely matches the noncommutativity structure of eqn. \eqref{eq:mainNCexampleYM}, and adding the extra term of $r^\prime$ precisely introduces the sign for $x^2$.

We believe this is a good indication that these deformations of $\ads$ provide gravitational dual descriptions of noncommutative $\mathcal{N}=4$ $\mathrm{U}(N)$ SYM,\footnote{Recall that the plane wave breaks translational symmetry of the center of mass of the branes.} noncommutative in the sense of eqn. \eqref{eq:mainNCexampleYM}. Note that the parameters in eqs. \eqref{eq:mainNCexampleYM} and \eqref{eq:mainNCexampleSTRING} are related by the effective string tension $T= \sqrt{\lambda}/{2\pi}$.\footnote{This matches our expectations from e.g. the Lunin-Maldacena(-Frolov) background, where we can ask what $r$ matrix would result in e.g. $\Phi_2 \Phi_3 \rightarrow e^{i \pi \gamma_1} \Phi_2 \Phi_3$ in the dual field theory. Given that the $\Phi_j$ transform in the fundamental of $\mathrm{SO}(6)$ and hence have charge $2i$ under the anti-hermitian $h_j$ of $\mathrm{SU}(4)$, we would expect $r = \pi T \gamma_1 (2i)^{-2} (2 h_2\wedge h_3)= -8^{-1} \sqrt{\lambda} \gamma_1 \epsilon^{1jk} h_j \wedge h_k$, precisely the $\hat{\gamma}_1$ term in the $r$ matrix \eqref{eq:Rmatrixbetadef} under the usual identification $\hat{\gamma}_i/\sqrt{\lambda}=\gamma_i$.} The equations themselves apply in the opposite domains of weakly coupled gauge theory and classical string theory respectively. Correspondingly, note that just like the canonical noncommutative case \cite{Hashimoto:1999ut,Maldacena:1999mh} the metric approaches that of undeformed $\ads$ at large $z$ (the infrared regime of the dual field theory), but differs for $z \sim \sqrt{a} R$ or $z \sim (a^2 R^4 (x_1^2 \pm x_2^2))^{1/3}$ depending on the region of space we consider, cf. eqs. \eqref{eq:mainNCexampleYM} and \eqref{eq:mainNCexampleSTRING}.

Now that we have further support for our conjecture in two nontrivial cases, let us look at some other possibly interesting noncommutative structures.

\section{$\kappa$-Minkowski space and $r$ matrices}
\label{sec:kappaminkowskiandsingularities}

Generalized $\kappa$-Minkowski space \cite{Majid:1994cy} corresponds to
\begin{equation}
\label{eq:kappaNCcomm}
[x^\mu \stackrel{\star}{,} x^\nu]= i \kappa^{-1} (a^\mu x^\nu-a^\nu x^\mu),
\end{equation}
with $|a_\mu a^\mu|=1$, where a timelike $a^\mu$ gives true $\kappa$-Minkowski space with $[x^0 \stackrel{\star}{,} x^j] = i\kappa^{-1} x^j$, but we could also consider spacelike or null $\kappa$-Minkowski space. These will prove instructive examples, having precisely the type of noncommutativity we might expect to be difficult to give a dual field theory interpretation.

These noncommutative spaces were originally considered as modules for the $\kappa$-Poincar\'e algebra \cite{Lukierski:1991pn,Lukierski:1992dt}. The associated $\kappa$-Poincar\'e $r$ matrices \cite{zakrzewski:1994,Borowiec:2013lca}
\begin{equation}
\label{eq:rkappa}
r= \kappa^{-1} m_{\mu\nu} \wedge p^\nu,
\end{equation}
where $\mu$ can be a timelike, null or spatial index, only solve the homogeneous CYBE in the null case, while the timelike and spacelike case satisfy the modified CYBE. Therefore, we will consider only the null $\kappa$-Poincar\'e $r$ matrix here. Still, $\kappa$-Minkowski space also arises as the module for a set of Drinfeld twists over the Poincar\'e-Weyl algebra. Here the various cases are obtained from a (quantum) Drinfeld twist based on the jordanian $r$ matrix \cite{Borowiec:2008uj,Dimitrijevic:2014dxa}
\begin{equation}
\label{eq:rkappaDp}
r = \kappa^{-1} D \wedge p_\mu.
\end{equation}
Let us now consider the associated deformations of $\ads$.

\subsection{$\kappa$ deformations of $\ads$}

The timelike $\kappa$ deformation of $\mathrm{AdS}_5$ corresponding to the timelike case of eqn. \eqref{eq:rkappaDp} is
\begin{equation}
\begin{aligned}
ds^2 & =\frac{z^2 (-dt^2+dr^2 + dz^2) -\kappa^{-2} (dr - r z^{-1} dz)^2}{z^4-\kappa^{-2}(z^2 + r^2)} + \frac{r^2 (d\theta^2 + \sin^2{\theta} d\phi^2)}{z^2} ,\\
B & = \kappa^{-1} \frac{z dz \wedge dt + r dr \wedge dt}{z^4-\kappa^{-2}(z^2 + r^2)},
\end{aligned}
\end{equation}
where we have introduced spherical coordinates $(r,\theta,\phi)$ on $\mathbb{R}^3$ and denote $x^0$ by $t$. The spacelike $\kappa$ deformation analogously yields
\begin{equation}
\begin{aligned}
ds^2 & =\frac{\nu^2(-d\beta^2 + \cosh^2{\beta} d \xi^2)}{z^2} + \frac{z^2 (dx_3^2+d\nu^2 + dz^2) +\kappa^{-2} (d\nu - \nu z^{-1} dz)^2}{z^4+\kappa^{-2}(z^2 + \nu^2)} ,\\
B & = \kappa^{-1} \frac{z dz \wedge dx^3 + \nu d\nu \wedge d x^3}{z^4+\kappa^{-2}(z^2 + \nu^2)},
\end{aligned}
\end{equation}
where we introduced `hyperbolic coordinates' on $\mathbb{R}^{1,2}$ via $t= \nu \sinh \beta$, $x^1 = \nu \cosh \beta \sin \xi$, and $x^2 = \nu\cosh \beta \cos \xi$. Finally the null deformation corresponding to null case of eqn. \eqref{eq:rkappaDp} is
\begin{align}
ds^2 & =\frac{z^2(-2 dx^+ dx^- + d\rho^2 + dz^2) - \kappa^{-2}z^{-2}((zdx^+ - x^+ dz)^2 + (x^+ d\rho - \rho dx^+)^2)}{z^4 -\kappa^{-2} (x^+)^2} + \frac{\rho^2 d\zeta^2}{z^2},\nonumber\\
B & = \kappa^{-1} \frac{-z dz \wedge dx^+ - x^+ dx^+ \wedge dx^- - \rho d \rho \wedge  dx^+}{z^4 -\kappa^{-2} (x^+)^2}, \label{eq:nullkappaAdS5}
\end{align}
where we took polar coordinates $\rho$ and $\zeta$ on the $(x_1,x_2)$ plane. Note that here the $z dz$ term in the B field is a total derivative. The background for the null $\kappa$-Poincar\'e case of eqn. \eqref{eq:rkappa} is given by dropping this total derivative in the $B$ field, and dropping the $ zdx^+$ term from the metric. The integrable sigma models associated to these spaces have no (manifest) supersymmetry.

The null deformed space \eqref{eq:nullkappaAdS5} is actually a special case of a background given in \cite{vanTongeren:2015soa}, based on the $r$ matrix
\begin{equation}
\label{eq:Rgen}
r = (a\, D - c \,m_{+-}) \wedge p_-,
\end{equation}
giving the null version eqn. \eqref{eq:rkappaDp} at $c=0$, and containing our main example of eqn. \eqref{eq:mainRmatrixexample} at $c=a$. We can express this $r$ matrix as a sum of the $r$ matrix for our main example and the abelian $r$ matrix $(D + m_{+-}) \wedge p_-$. On its own this abelian $r$ matrix corresponds to a (formal) TsT transformation in $(y,x^-)$, where $y$ is the coordinate associated with the boost-dilation (Schr\"odinger dilation) $x^+ \rightarrow e^{2\alpha} x^+$, $z\rightarrow e^\alpha z$, $\rho\rightarrow e^\alpha \rho$ generated by $D + m_{+-}$ \cite{vanTongeren:2015soa}. Since $y$ and $x^-$ are still isometry coordinates of the deformed geometry \eqref{eq:mainJordanianDefAdS5}, we can append this TsT transformation to the sequence of the previous section, which  indeed gives the general background of \cite{vanTongeren:2015soa} and in particular the null $\kappa$-Minkowski deformed $\mathrm{AdS}_5$ of eqn. \eqref{eq:nullkappaAdS5}, which is thereby embedded in supergravity.\footnote{At the algebraic level the corresponding picture is that $(D + m_{+-}) \wedge p_-$ is subordinate \cite{Tolstoy:2008zz} to $(D - m_{+-}) \wedge p_-$, so that their sum is a solution of the CYBE, and the total twist factorizes into a product of an abelian and a jordanian piece.} We have not investigated this for the timelike and spacelike cases, or for the $\kappa$-Poincar\'e version of the null case, though it is very similar and may well be accounted for with minor modifications. Furthermore, since this TsT transformation involves isometries of $\mathrm{AdS}_5$ that are not isometries of the brane background, we cannot directly use this trick there.

At large $z$ these spaces approach undeformed $\ads$, but differences become apparent as we decrease, and in particular the timelike and null $\kappa$-Minkowski deformations of $\ads$ become very different; they are singular. In the timelike case we encounter a singularity at $z^2 = (\kappa^{-2} + \sqrt{\kappa^{-4} + \kappa^{-2} r^2})/2$, while for the null case we do so at $z^2 = \kappa^{-1} |x^+|$. As discussed in section \ref{subsec:mainconjecture}, the presence of singularities makes it unclear whether we can give a dual field theory interpretation to the timelike and null $\kappa$-Minkowski cases.\footnote{ While it might appear tempting to link the corresponding singularities in the $B$ fields above to the critical values in the electric field for the canonical temporal-spatial noncommutative model mentioned in section \ref{subsec:mainconjecture}, we should keep in mind that at this stage we cannot assume these spaces are actually gravity duals of something, and even if they were, here we are dealing with a `near horizon', closed string quantity.}

\subsection{A gravity dual for SYM on spacelike $\kappa$-Minkowski space?}

The spacelike case would suggestively pass the test of regularity, and we are tempted to conjecture that it is the gravity dual of SYM on spacelike $\kappa$-Minkowski space. Of course, this comes with disclaimers: we have not embedded this space in supergravity, or consequently attempted to find a brane picture for the associated noncommutativity, and we may well run into difficulties besides regularity, also more directly on the field theory side. It would be interesting to investigate this further. The regularity of the spacelike case contrasted with the null and timelike cases is clearly related to the fact that when time appears in the noncommutative structure, it does so homogeneously, but we have no deeper interpretation to offer at this time. Let us emphasize again that we are considering $\kappa$-Minkowski space in a twisted setting, so that the associated symmetry structure is not that of the $\kappa$-Poincar\'e group, except in one of our two null cases.

\section{Conclusions}
\label{sec:conclusions}

Homogeneous Yang-Baxter deformations of the $\ads$ superstring have a deformed symmetry algebra of a type that can occur in $\mathcal{N}=4$ SYM if defined on an appropriate noncommutative space. We conjectured that these two classes of deformed theories are generically gauge/gravity duals of one another, though subtleties may arise in the presence of singularities. For abelian deformations this conjecture gives a unified perspective on a list of established dualities, while it was entirely new for jordanian ones. The gravitational perspective this offers should be able to provide some insights into twist-noncommutative field theory in general. Of course, any of these noncommutative versions of SYM should be integrable in the planar limit,\footnote{Still, it is the integrability preserving nature of these deformations as opposed to their inherent integrability that appears to be of main relevance for our story; the twists arose from nonlocal gauge transformations relating deformed currents to the undeformed one, preserving flatness, hence integrability.} though how this would precisely manifest itself is not immediately clear.

We checked our conjecture in the jordanian case with two examples with noncommutativity of the type
\begin{equation}
\label{eq:NCofexamples}
[x^-\stackrel{\star}{,}x^j] = (\pm) i a \, x^j, \hspace{10pt} j=1,2,
\end{equation}
by finding a matching brane picture. Let us emphasize that this noncommutativity is different from ``isometric'' ones obtained by TsT transformations (abelian deformations), such as
\begin{equation*}
[x^-\stackrel{\star}{,}x^1] = i c \,x^2, \hspace{20pt}
[x^-\stackrel{\star}{,}x^2] = -i c \,x^1
\end{equation*}
for $r = c \, m_{12}\wedge p_-$. Given the concrete AdS/CFT interpretation we have found for these two jordanian deformations, it may be interesting to do various `classic' AdS/CFT computations regarding correlation functions and Wilson loops in these deformed dual geometries, (as well as) to understand the `boundary' geometry of these deformed spaces in relation to the noncommutativity \eqref{eq:NCofexamples}.

We also studied $\kappa$-Minkowski--related deformations of $\ads$, where in particular the spacelike case shows interesting features. Based on the regularity of the spacelike deformation, though lacking direct support in the form of a brane picture, we conjectured that this deformation may be dual to the corresponding noncommutative version of SYM. Finding further support for, or subtleties with this conjecture is certainly an important direction for future investigation.\footnote{Though outside the scope of the present paper, regarding these $\kappa$ deformations it is interesting to recall that the $\kappa$-Poincar\'e group was originally obtained as a contraction limit of $\mathrm{SO}_q(3,2)$, leading us to wonder whether a similar contraction limit of $\mathrm{SO}_q(4,2)$ and $\mathrm{PSU}_q(2,2|4)$ could have an interesting implementation in the quantum deformed $\ads$ sigma model.}

The fact that the second of our main examples as well as the (spacelike) $\kappa$-Minkowski case are not supersymmetric may lead to subtleties. Let us elaborate on this point by example. As we already implicitly saw above, there is a ``natural'' deformation of SYM for the three parameter generalization of the Lunin-Maldacena background \cite{Frolov:2005dj,Frolov:2005iq} (which our construction would indeed exactly suggest). This ``$\gamma_i$ deformed'' SYM has no supersymmetry, and its classical conformal symmetry is broken at the quantum level even in the planar limit \cite{Fokken:2013aea}, leading to questions about its AdS/CFT interpretation. However, without supersymmetry some of the string modes may become tachyonic at the quantum level and lead to a deformation of $\mathrm{AdS}_5$ as the analogue of breaking conformal invariance. Hence, despite lacking a clean AdS/CFT interpretation, there may well be a duality, and correspondingly it appears to be possible to match spectra between the two theories at least in the planar limit, for a large class of states, see \cite{Fokken:2014soa,vanTongeren:2013gva} and references therein. Similar subtleties may arise for nonsupersymmetric jordanian deformations.

Moreover, regarding supersymmetry, while we presently only considered bosonic generators, we could actually try to `supersymmetrize' our $r$ matrices by adding fermionic bilinears, see e.g. \cite{Borowiec:2008se}. Provided such $r$ matrices exist here, we suppose they would directly modify the RR fluxes of the background, affecting the supersymmetry properties of the background without affecting its metric and (bosonic) B field. If purely fermionic abelian cases exist they presumably correspond to some sort of fermionic analogue of a TsT transformation, based on fermionic T duality \cite{Berkovits:2008ic,Beisert:2008iq}. Note that the simple $r$ matrix based (super)symmetry analysis \cite{vanTongeren:2015soa} would change correspondingly.

It would also be interesting to understand whether singular integrable backgrounds can be given a meaningful interpretation in general, even just as strings, given their integrability. The free string picture is rather clear for the already problematic abelian $p_0 \wedge p_1$ deformation at least, but less so for generic (jordanian) ones.

In general it would be great to understand whether all homogeneous Yang-Baxter deformations of $\ads$ can be embedded in supergravity. At the abelian level there is a clear link to TsT transformations, which can perhaps be proven. For jordanian deformations, however, there is presently no clear relation to supergravity solution generating techniques. Are they always embeddable, and if so do they correspond to a new solution generating technique, or can they be fully understood in terms of known ones? Of course, in trying to interpret these deformations in terms of supergravity it may be helpful to have an explicit classification of solutions to the Yang-Baxter equation over $\mathfrak{psu}(2,2|4)$. No such classification exists for generic simple Lie algebras, but solutions over the Poincar\'e algebra have been classified \cite{zakrzewski:1997}. In that particular case the analysis was likely facilitated by the semidirect product structure of the Poincar\'e algebra and the isomorphism of the Lorentz algebra to $\mathfrak{sl}(2, \mathbb{C})$ however. We have investigated some solutions of \cite{zakrzewski:1997}, the interested reader can find one such Poincar\'e based (singular) deformation of $\ads$ in appendix \ref{app:example}. Finally, since much of the tools based on integrability in AdS/CFT rely on an exact S matrix approach, it may be fruitful to look for further light cone gauge compatible deformations of global anti-de Sitter space, or ways to adapt this approach.

\section*{Acknowledgements}

I would like to thank B. Hoare and A. Pacho{\l} for insightful discussions, and G. Arutyunov and A. Tseytlin for comments on the paper. ST is supported by LT. This work was supported by the Einstein Foundation Berlin in the framework of the research project "Gravitation and High Energy Physics", and acknowledges further support from the People Programme (Marie Curie Actions) of the European Union's Seventh Framework Programme FP7/2007-2013/ under REA Grant Agreement No. 317089 (GATIS).

\appendix

\section{Algebra, coordinates, and the sigma model action}
\label{app:algebraetc}

\subsection{Matrix realization of $\mathfrak{su}(2,2)\oplus\mathfrak{su}(4)$}

In this paper we are mainly concerned with the bosonic subalgebra $\mathfrak{su}(2,2)\oplus\mathfrak{su}(4)$ of $\mathfrak{\psu}(2,2|4)$. For details on the material presented here, as well as its supersymmetric extension, we refer to the pedagogical review \cite{Arutyunov:2009ga} whose conventions we follow. We only briefly list needed facts, beginning with the $\gamma$ matrices
\begin{equation}
\begin{aligned}
&\gamma^0 = i \sigma_3 \otimes \sigma_0,  &\gamma^1 = \sigma_2 \otimes \sigma_2, &&\gamma^2 = -\sigma_2 \otimes \sigma_1, \\ &\gamma^3 = \sigma_1 \otimes \sigma_0, &\gamma^4 = \sigma_2 \otimes \sigma_3, &&\gamma^5 = -i \gamma^0,
\end{aligned}
\end{equation}
where $\sigma_0 = 1_{2\times2}$ and the remaining $\sigma_i$ are the Pauli matrices. With these matrices the generators of $\mathfrak{so}(4,1)$ in the spinor representation are given by $m^{ij} = \frac{1}{4} [\gamma^i,\gamma^j]$ where the indices run from zero to four, while for $\mathfrak{so}(5)$ we can give the same construction with indices running from one to five. The algebra $\mathfrak{su}(2,2)$ is spanned by these generators of $\mathfrak{so}(4,1)$ together with the $\gamma^i$ for $i=0,\ldots,4$, while $\mathfrak{su}(4)$ is spanned by the combination of $\mathfrak{so}(5)$ and $i\gamma^j$ for $j=1,\ldots,5$. These generators satisfy
\begin{equation}
m^\dagger \gamma^5 + \gamma^5 m = 0
\end{equation}
for $m\in \mathfrak{su}(2,2)$, and
\begin{equation}
n^\dagger+ n = 0
\end{equation}
for $n \in \mathfrak{su}(4)$. This means that we are dealing with the canonical group metric $\gamma^5 = \mathrm{diag}(1,1,-1,-1)$ for $\mathrm{SU}(2,2)$, and that $e^{\alpha n}$ and $e^{\alpha m}$ give group elements for real $\alpha$.

The generator $\Omega$ of the $\mathbb{Z}_4$ automorphism of $\mathfrak{\psu}(2,2|4)$ acts on these bosonic subalgebras as
\begin{equation}
\label{eq:Z4gendef}
\Omega(m) = -K m^t K,
\end{equation}
where $K=-\gamma^2\gamma^4$, which leaves the subalgebras $\mathfrak{so}(4,1)$ and $\mathfrak{so}(5)$ invariant.

We denote the Cartan generators of $\mathfrak{su}(4)$ by $h_j$, $j=1,2,3$, taken to be
\begin{equation}
h_1 = \mathrm{diag}(i,i,-i,-i), \hspace{15pt} h_2 = \mathrm{diag}(i,-i,i,-i), \hspace{15pt} h_3 = \mathrm{diag}(i,-i,-i,i).
\end{equation}
Our group parametrization below will use the Poincar\'e translation generators
\begin{equation}
p^\mu = \frac{1}{2}(\gamma^\mu - \gamma^\mu \gamma^4),
\end{equation}
as well as the special conformal generators
\begin{equation}
k^\mu = \frac{1}{2}(\gamma^\mu + \gamma^\mu \gamma^4).
\end{equation}
Both the $p$ and the $k$ are nilpotent (as matrices)
\begin{equation}
p^\mu p^\nu = k^\mu k^\nu =0.
\end{equation}

\subsection{Vector field realization of $\mathfrak{su}(2,2)$}
\label{app:su22vec}

We can represent the conformal algebra in terms of differential operators as
\begin{equation}
\begin{aligned}
M_{\mu\nu} & = x_\mu \partial_\nu - x_\nu \partial_\mu, \hspace{20pt} K_\mu  = x_\alpha x^\alpha \partial_\mu -2 x_\mu x^\nu \partial_\nu \\
P_\mu & = \partial_\mu \hspace{20pt} D = - x^\mu \partial_\mu,
\end{aligned}
\end{equation}
which satisfy
\begin{equation}
\begin{aligned}
[M_{\mu\nu}, P_\rho] &  = \eta_{\nu\rho} P_\mu - \eta_{\mu\rho} P_\nu, \hspace{20pt} [M_{\mu\nu}, K_\rho]  = \eta_{\nu\rho} K_\mu- \eta_{\mu\rho} K_\nu,\\
[M_{\mu\nu}, D] & = 0, \hspace{20pt} [D,P_\mu]=P_\mu, \hspace{20pt} [D,K_\mu]=-K_\mu,\\
[P_\mu,K_\nu] & =2 M_{\mu\nu} + 2 \eta_{\mu\nu} D, \hspace{20pt}
[M_{\mu\nu},M_{\rho\sigma}] = \eta_{\mu\rho} M_{\nu\sigma} + \mbox{perms.}
\end{aligned}
\end{equation}
Note that we work with anti-Hermitian generators. These generators realize the matrix generators $m(M)$, $p(P)$ and $k(K)$ of the previous section as vector fields, with $D$ corresponding to $\tfrac{1}{2}\gamma_4$. In the main text we refer to $\tfrac{1}{2}\gamma_4$ as $D$ directly, and use $m$, $p$, and $k$ to denote the generators in any appropriate realization.

\subsection{Bosonic sigma model action}

The concrete metrics and B fields in this paper can be read off from the bosonic part of the deformed sigma model. In these bosonic models we work with the coset representative
\begin{equation}
g = \left(\begin{array}{cc}g_a & 0 \\ 0 & g_s \end{array}\right),
\end{equation}
with
\begin{equation}
g_a = e^{x_\mu p^\mu} e^{ \log z D}=(1+x_\mu p^\mu) e^{ \log z D},
\end{equation}
and for completeness
\begin{equation}
g_s = e^{\phi^i h_i} e^{- \frac{\xi}{2} \gamma^1 \gamma^3} e^{\frac{i}{2} \arcsin{r} \gamma^1}.
\end{equation}
Substituting this group element in the undeformed version of eqn. \eqref{eq:defaction}, gives the (bosonic) string action of $\ads$ in Poincar\'e coordinates, i.e. an action of the form
\begin{equation}
\label{eq:genericstringaction}
S=-\tfrac{T}{2}\int{\rm d}\tau{\rm d}\sigma \, \left(g_{\mathsc{mn}}\, dx^\mathsc{m} dx^\mathsc{n} - B_{\mathsc{mn}}\, dx^\mathsc{m} \hspace{-3pt} \wedge dx^\mathsc{n}\right) ,
\end{equation}
where now $g_{\mathsc{mn}}$ is the metric of $\ads$ and the B field is zero. A coset representative that gives global AdS can be found in e.g. \cite{Arutyunov:2009ga} or \cite{Arutyunov:2013ega}. The metric and B field corresponding to the deformed action \eqref{eq:defaction} can be readily extracted following \cite{Arutyunov:2013ega,Kawaguchi:2014fca}. We need to construct the current $J$, which is defined as in the main text via
\begin{equation}
J=(1-\eta R_g \circ d)^{-1}(A)
\end{equation}
The operator $1-\eta R_g \circ d$ can be inverted, at least for general values of the group coordinates, and generally perturbatively in $\eta$ \cite{Arutyunov:2013ega,Delduc:2014kha}. As grades one and three are fermionic, in practice we only need $J^{(2)}$ for the bosonic action, where we have
\begin{equation}
\label{eq:J2eqn}
P_2(A)= A^{(2)}=P_2\left((1-2\eta R_g \circ P_2)(J)\right)= P_2\left((1-2\eta R_g)(J^{(2)})\right).
\end{equation}
Using a computer this equation can be readily solved for $J^{(2)}$, and substituted in the action, from which the metric and B field can be read off by comparing to eqn. \eqref{eq:genericstringaction} above. To be clear, cf. eqn. \eqref{eq:Z4gendef} we have $P_2(X)=\tfrac{1}{2}(X-\Omega(X))$. In practice it is useful to expand eqn. \eqref{eq:J2eqn} over the basis of the grade two components, rather than work directly with the matrix realization of $\mathfrak{su}(2,2|4)$.

\subsection{$\mathrm{S}^5$ as $\mathrm{S}^1$ over $\mathbb{CP}^2$}
\label{app:fibration}

To describe $\mathrm{S}^5$ as an $\mathrm{S}^1$ fibration over $\mathbb{CP}^2$ we can embed a five sphere of radius one  in $\mathbb{R}^6$ with Cartesian coordinates $y_j$ via
\begin{equation}
\begin{aligned}
y_1+i y_2 & =e^{i \chi} \sin{\mu} \cos\tfrac{\theta}{2} e^{\tfrac{i}{2}(\psi+\phi)},\\
y_3+i y_4 & =e^{i \chi} \sin{\mu} \sin\tfrac{\theta}{2} e^{\tfrac{i}{2}(\psi-\phi)},\\
y_5+i y_6 & =e^{i \chi} \cos{\mu}.
\end{aligned}
\end{equation}
This results in the metric
\begin{equation}
ds^2 = (d\chi + \omega)^2 + d\mu^2 +\sin^2\mu(s_1^2 + s_2^2 + \cos^2 \mu\, s_3^2),
\end{equation}
where $\omega = \sin^2 \mu \, s_3$, and
\begin{equation}
\begin{aligned}
s_1& =\frac{1}{2}(\cos \psi d \theta + \sin \psi \sin \theta d\phi),\\
s_2& =\frac{1}{2}(\sin \psi d \theta - \cos \psi \sin \theta d\phi),\\
s_3& =\frac{1}{2}(d\psi + \cos \theta d\phi).
\end{aligned}
\end{equation}
Note that
\begin{equation}
d\chi + \omega = y_1 d y_2 - y_2 d y_1 + y_3 d y_4 - y_4 d y_3 + y_5 d y_6 - y_6 d y_5,
\end{equation}
which for $\mathbb{R}^6$ in terms of hyperspherical coordinates in the fibered sense really means
\begin{equation}
r^2(d\chi + \omega) =x_1 d x_2 - x_2 d x_1 +  x_3 d x_4 - x_4 d x_3 + x_5 d x_6 - x_6 d x_5,
\end{equation}
where $r$ is the radial coordinate and the $x_i$ are (unconstrained) Cartesian coordinates. Note that in the main text the indices are shifted by two since $x_1$ and $x_2$ are already used.

\subsection{A Poincar\'e based deformation of $\ads$}
\label{app:example}

To give a (random) example of the sort of deformations one can get by considering Poincar\'e $r$ matrices we consider the jordanian $r$ matrix \cite{zakrzewski:1997,Tolstoy:2008zz}
\begin{equation}
r = - a m_{02} \wedge (m_{12}+m_{01}),
\end{equation}
would naively correspond to noncommutativity of the type
\begin{equation}
[x^{m}\stackrel{\star}{,}x^{n}]= i a \epsilon^{mn}_{\,\,\,\,\,\,\,\, p}x^{p} (x^0+x^2),
\end{equation}
where indices $m$,$n$ and $p$ run from zero to two and $\epsilon$ is totally antisymmetric with $\epsilon^{012}=1=-\epsilon_0^{\,\,\,12}$. The corresponding deformation of $\ads$ is given by
\begin{equation}
\begin{aligned}
ds^2 & =\frac{z^2 \nu^2 (-d\beta^2 + \cosh^2\beta d\xi^2)}{z^4 - a^2(\sinh \beta + \cos \xi \cosh \beta)^2 \nu^4} + \frac{d\nu^2 + dx_3^2 + dz^2}{z^2} ,\\
B & = a \frac{(\sinh \beta + \cos \xi \cosh \beta)\cosh \beta \nu^4}{z^4 - a^2(\sinh \beta + \cos \xi \cosh \beta)^2 \nu^4} d \xi\wedge d\beta ,
\end{aligned}
\end{equation}
using the hyperbolic coordinates of the spacelike $\kappa$-Minkowski deformation in the main text. We chose this example out of many, including abelian ones, where the deformation parameter has dimensions of inverse length squared. This space is singular.

\bibliographystyle{JHEP}

\bibliography{Stijnsbibfile}

\providecommand{\href}[2]{#2}\begingroup\raggedright\begin{thebibliography}{100}

\bibitem{Maldacena:1997re}
J.~M. Maldacena, {\it {The Large N limit of superconformal field theories and
  supergravity}},  {\em Int. J. Theor. Phys.} {\bf 38} (1999) 1113--1133,
  [\href{http://arxiv.org/abs/hep-th/9711200}{{\tt hep-th/9711200}}]. [Adv.
  Theor. Math. Phys.2,231(1998)].

\bibitem{Arutyunov:2009ga}
G.~Arutyunov and S.~Frolov, {\it {Foundations of the $\ads$ Superstring. Part
  I}},  {\em J.Phys.} {\bf A42} (2009) 254003,
  [\href{http://arxiv.org/abs/0901.4937}{{\tt arXiv:0901.4937}}].

\bibitem{Beisert:2010jr}
N.~Beisert, C.~Ahn, L.~F. Alday, Z.~Bajnok, J.~M. Drummond, et~al., {\it
  {Review of AdS/CFT Integrability: An Overview}},  {\em Lett.Math.Phys.} {\bf
  99} (2012) 3--32, [\href{http://arxiv.org/abs/1012.3982}{{\tt
  arXiv:1012.3982}}].

\bibitem{Lunin:2005jy}
O.~Lunin and J.~M. Maldacena, {\it {Deforming field theories with $U(1)\times
  U(1)$ global symmetry and their gravity duals}},  {\em JHEP} {\bf 0505}
  (2005) 033, [\href{http://arxiv.org/abs/hep-th/0502086}{{\tt
  hep-th/0502086}}].

\bibitem{Frolov:2005ty}
S.~Frolov, R.~Roiban, and A.~A. Tseytlin, {\it {Gauge-string duality for
  superconformal deformations of N=4 super Yang-Mills theory}},  {\em JHEP}
  {\bf 0507} (2005) 045, [\href{http://arxiv.org/abs/hep-th/0503192}{{\tt
  hep-th/0503192}}].

\bibitem{Frolov:2005dj}
S.~Frolov, {\it {Lax pair for strings in Lunin-Maldacena background}},  {\em
  JHEP} {\bf 0505} (2005) 069, [\href{http://arxiv.org/abs/hep-th/0503201}{{\tt
  hep-th/0503201}}].

\bibitem{Delduc:2013qra}
F.~Delduc, M.~Magro, and B.~Vicedo, {\it {An integrable deformation of the
  $\ads$ superstring action}},  {\em Phys.Rev.Lett.} {\bf 112} (2014) 051601,
  [\href{http://arxiv.org/abs/1309.5850}{{\tt arXiv:1309.5850}}].

\bibitem{Klimcik:2002zj}
C.~Klimcik, {\it {Yang-Baxter sigma models and dS/AdS T duality}},  {\em JHEP}
  {\bf 0212} (2002) 051, [\href{http://arxiv.org/abs/hep-th/0210095}{{\tt
  hep-th/0210095}}].

\bibitem{Klimcik:2008eq}
C.~Klimcik, {\it {On integrability of the Yang-Baxter sigma-model}},  {\em
  J.Math.Phys.} {\bf 50} (2009) 043508,
  [\href{http://arxiv.org/abs/0802.3518}{{\tt arXiv:0802.3518}}].

\bibitem{Delduc:2013fga}
F.~Delduc, M.~Magro, and B.~Vicedo, {\it {On classical $q$-deformations of
  integrable sigma-models}},  {\em JHEP} {\bf 1311} (2013) 192,
  [\href{http://arxiv.org/abs/1308.3581}{{\tt arXiv:1308.3581}}].

\bibitem{Arutyunov:2013ega}
G.~Arutyunov, R.~Borsato, and S.~Frolov, {\it {S-matrix for strings on
  $\eta$-deformed $AdS_{5} \times S^5$}},  {\em JHEP} {\bf 1404} (2014) 002,
  [\href{http://arxiv.org/abs/1312.3542}{{\tt arXiv:1312.3542}}].

\bibitem{Delduc:2014kha}
F.~Delduc, M.~Magro, and B.~Vicedo, {\it {Derivation of the action and
  symmetries of the $q$-deformed $\ads$ superstring}},  {\em JHEP} {\bf 1410}
  (2014) 132, [\href{http://arxiv.org/abs/1406.6286}{{\tt arXiv:1406.6286}}].

\bibitem{Lunin:2014tsa}
O.~Lunin, R.~Roiban, and A.~Tseytlin, {\it {Supergravity backgrounds for
  deformations of AdS$_{n} \times \rm{S}^n$ supercoset string models}},  {\em
  Nucl.Phys.} {\bf B891} (2015) 106--127,
  [\href{http://arxiv.org/abs/1411.1066}{{\tt arXiv:1411.1066}}].

\bibitem{Arutynov:2014ota}
G.~Arutyunov, M.~de~Leeuw, and S.~J. van Tongeren, {\it {The exact spectrum and
  mirror duality of the $(\ads)_\eta$ superstring}},  {\em Theor.Math.Phys.}
  {\bf 182} (2015), no.~1 23--51, [\href{http://arxiv.org/abs/1403.6104}{{\tt
  arXiv:1403.6104}}].

\bibitem{Arutyunov:2014cra}
G.~Arutyunov and S.~J. van Tongeren, {\it {The $\mathrm{AdS}_5 \times
  \mathrm{S}^5$ mirror model as a string}},  {\em Phys.Rev.Lett.} {\bf 113}
  (2014) 261605, [\href{http://arxiv.org/abs/1406.2304}{{\tt
  arXiv:1406.2304}}].

\bibitem{Arutyunov:2014jfa}
G.~Arutyunov and S.~J. van Tongeren, {\it {Double Wick rotating Green-Schwarz
  strings}},  {\em JHEP} {\bf 1505} (2015) 027,
  [\href{http://arxiv.org/abs/1412.5137}{{\tt arXiv:1412.5137}}].

\bibitem{Hoare:2014pna}
B.~Hoare, R.~Roiban, and A.~Tseytlin, {\it {On deformations of $AdS_n$ x $S^n$
  supercosets}},  {\em JHEP} {\bf 1406} (2014) 002,
  [\href{http://arxiv.org/abs/1403.5517}{{\tt arXiv:1403.5517}}].

\bibitem{Sfetsos:2013wia}
K.~Sfetsos, {\it {Integrable interpolations: From exact CFTs to non-Abelian
  T-duals}},  {\em Nucl.Phys.} {\bf B880} (2014) 225--246,
  [\href{http://arxiv.org/abs/1312.4560}{{\tt arXiv:1312.4560}}].

\bibitem{Hollowood:2014qma}
T.~J. Hollowood, J.~L. Miramontes, and D.~M. Schmidtt, {\it {An Integrable
  Deformation of the $\ads$ Superstring}},  {\em J.Phys.} {\bf A47} (2014),
  no.~49 495402, [\href{http://arxiv.org/abs/1409.1538}{{\tt
  arXiv:1409.1538}}].

\bibitem{Demulder:2015lva}
S.~Demulder, K.~Sfetsos, and D.~C. Thompson, {\it {Integrable
  $\lambda$-deformations: Squashing Coset CFTs and $AdS_5\times S^5$}},
  \href{http://arxiv.org/abs/1504.02781}{{\tt arXiv:1504.02781}}.

\bibitem{Sfetsos:2014cea}
K.~Sfetsos and D.~C. Thompson, {\it {Spacetimes for $\lambda$-deformations}},
  \href{http://arxiv.org/abs/1410.1886}{{\tt arXiv:1410.1886}}.

\bibitem{Vicedo:2015pna}
B.~Vicedo, {\it {Deformed integrable $\sigma$-models, classical $R$-matrices
  and classical exchange algebra on Drinfel'd doubles}},
  \href{http://arxiv.org/abs/1504.06303}{{\tt arXiv:1504.06303}}.

\bibitem{Hoare:2015gda}
B.~Hoare and A.~A. Tseytlin, {\it {On integrable deformations of superstring
  sigma models related to AdS$_n$ $\times$ S$^n$ supercosets}},  {\em Nucl.
  Phys.} {\bf B897} (2015) 448--478,
  [\href{http://arxiv.org/abs/1504.07213}{{\tt arXiv:1504.07213}}].

\bibitem{Balog:1993es}
J.~Balog, P.~Forgacs, Z.~Horvath, and L.~Palla, {\it {A New family of SU(2)
  symmetric integrable sigma models}},  {\em Phys.Lett.} {\bf B324} (1994)
  403--408, [\href{http://arxiv.org/abs/hep-th/9307030}{{\tt hep-th/9307030}}].

\bibitem{Kawaguchi:2014qwa}
I.~Kawaguchi, T.~Matsumoto, and K.~Yoshida, {\it {Jordanian deformations of the
  $\ads$ superstring}},  {\em JHEP} {\bf 1404} (2014) 153,
  [\href{http://arxiv.org/abs/1401.4855}{{\tt arXiv:1401.4855}}].

\bibitem{Matsumoto:2014nra}
T.~Matsumoto and K.~Yoshida, {\it {Lunin-Maldacena backgrounds from the
  classical Yang-Baxter equation - towards the gravity/CYBE correspondence}},
  {\em JHEP} {\bf 1406} (2014) 135, [\href{http://arxiv.org/abs/1404.1838}{{\tt
  arXiv:1404.1838}}].

\bibitem{Matsumoto:2014gwa}
T.~Matsumoto and K.~Yoshida, {\it {Integrability of classical strings dual for
  noncommutative gauge theories}},  {\em JHEP} {\bf 1406} (2014) 163,
  [\href{http://arxiv.org/abs/1404.3657}{{\tt arXiv:1404.3657}}].

\bibitem{vanTongeren:2015soa}
S.~J. van Tongeren, {\it {On classical Yang-Baxter based deformations of the
  $\ads$ superstring}},  {\em JHEP} {\bf 06} (2015) 048,
  [\href{http://arxiv.org/abs/1504.05516}{{\tt arXiv:1504.05516}}].

\bibitem{Kawaguchi:2014fca}
I.~Kawaguchi, T.~Matsumoto, and K.~Yoshida, {\it {A Jordanian deformation of
  AdS space in type IIB supergravity}},  {\em JHEP} {\bf 1406} (2014) 146,
  [\href{http://arxiv.org/abs/1402.6147}{{\tt arXiv:1402.6147}}].

\bibitem{Matsumoto:2014ubv}
T.~Matsumoto and K.~Yoshida, {\it {Yang-Baxter deformations and string
  dualities}},  {\em JHEP} {\bf 1503} (2015) 137,
  [\href{http://arxiv.org/abs/1412.3658}{{\tt arXiv:1412.3658}}].

\bibitem{DrinfeldTwistRef}
V.~Drinfeld, {\it {On constant quasi-classical solutions of the Yang-Baxter
  quantum equation}},  {\em Sov. Math. Dokl.} {\bf 28} (1983) 667.

\bibitem{Douglas:2001ba}
M.~R. Douglas and N.~A. Nekrasov, {\it {Noncommutative field theory}},  {\em
  Rev.Mod.Phys.} {\bf 73} (2001) 977--1029,
  [\href{http://arxiv.org/abs/hep-th/0106048}{{\tt hep-th/0106048}}].

\bibitem{Szabo:2001kg}
R.~J. Szabo, {\it {Quantum field theory on noncommutative spaces}},  {\em
  Phys.Rept.} {\bf 378} (2003) 207--299,
  [\href{http://arxiv.org/abs/hep-th/0109162}{{\tt hep-th/0109162}}].

\bibitem{Chaichian:2004za}
M.~Chaichian, P.~Kulish, K.~Nishijima, and A.~Tureanu, {\it {On a
  Lorentz-invariant interpretation of noncommutative space-time and its
  implications on noncommutative QFT}},  {\em Phys.Lett.} {\bf B604} (2004)
  98--102, [\href{http://arxiv.org/abs/hep-th/0408069}{{\tt hep-th/0408069}}].

\bibitem{Chaichian:2004yh}
M.~Chaichian, P.~Presnajder, and A.~Tureanu, {\it {New concept of relativistic
  invariance in NC space-time: Twisted Poincare symmetry and its
  implications}},  {\em Phys.Rev.Lett.} {\bf 94} (2005) 151602,
  [\href{http://arxiv.org/abs/hep-th/0409096}{{\tt hep-th/0409096}}].

\bibitem{Aschieri:2005zs}
P.~Aschieri, M.~Dimitrijevic, F.~Meyer, and J.~Wess, {\it {Noncommutative
  geometry and gravity}},  {\em Class.Quant.Grav.} {\bf 23} (2006) 1883--1912,
  [\href{http://arxiv.org/abs/hep-th/0510059}{{\tt hep-th/0510059}}].

\bibitem{Aschieri:2006ye}
P.~Aschieri, M.~Dimitrijevic, F.~Meyer, S.~Schraml, and J.~Wess, {\it {Twisted
  gauge theories}},  {\em Lett.Math.Phys.} {\bf 78} (2006) 61--71,
  [\href{http://arxiv.org/abs/hep-th/0603024}{{\tt hep-th/0603024}}].

\bibitem{Aschieri:2007sq}
P.~Aschieri, F.~Lizzi, and P.~Vitale, {\it {Twisting all the way: From
  Classical Mechanics to Quantum Fields}},  {\em Phys.Rev.} {\bf D77} (2008)
  025037, [\href{http://arxiv.org/abs/0708.3002}{{\tt arXiv:0708.3002}}].

\bibitem{Szabo:2006wx}
R.~J. Szabo, {\it {Symmetry, gravity and noncommutativity}},  {\em
  Class.Quant.Grav.} {\bf 23} (2006) R199--R242,
  [\href{http://arxiv.org/abs/hep-th/0606233}{{\tt hep-th/0606233}}].

\bibitem{Dimitrijevic:2014dxa}
M.~Dimitrijevic, L.~Jonke, and A.~Pacho\l{}, {\it {Gauge Theory on Twisted
  $\kappa$-Minkowski: Old Problems and Possible Solutions}},  {\em SIGMA} {\bf
  10} (2014) 063, [\href{http://arxiv.org/abs/1403.1857}{{\tt
  arXiv:1403.1857}}].

\bibitem{Seiberg:1999vs}
N.~Seiberg and E.~Witten, {\it {String theory and noncommutative geometry}},
  {\em JHEP} {\bf 9909} (1999) 032,
  [\href{http://arxiv.org/abs/hep-th/9908142}{{\tt hep-th/9908142}}].

\bibitem{Madore:2000en}
J.~Madore, S.~Schraml, P.~Schupp, and J.~Wess, {\it {Gauge theory on
  noncommutative spaces}},  {\em Eur.Phys.J.} {\bf C16} (2000) 161--167,
  [\href{http://arxiv.org/abs/hep-th/0001203}{{\tt hep-th/0001203}}].

\bibitem{Jurco:2001rq}
B.~Jurco, L.~Moller, S.~Schraml, P.~Schupp, and J.~Wess, {\it {Construction of
  nonAbelian gauge theories on noncommutative spaces}},  {\em Eur.Phys.J.} {\bf
  C21} (2001) 383--388, [\href{http://arxiv.org/abs/hep-th/0104153}{{\tt
  hep-th/0104153}}].

\bibitem{Connes:1997cr}
A.~Connes, M.~R. Douglas, and A.~S. Schwarz, {\it {Noncommutative geometry and
  matrix theory: Compactification on tori}},  {\em JHEP} {\bf 9802} (1998) 003,
  [\href{http://arxiv.org/abs/hep-th/9711162}{{\tt hep-th/9711162}}].

\bibitem{Douglas:1997fm}
M.~R. Douglas and C.~M. Hull, {\it {D-branes and the noncommutative torus}},
  {\em JHEP} {\bf 9802} (1998) 008,
  [\href{http://arxiv.org/abs/hep-th/9711165}{{\tt hep-th/9711165}}].

\bibitem{Chu:1998qz}
C.-S. Chu and P.-M. Ho, {\it {Noncommutative open string and D-brane}},  {\em
  Nucl.Phys.} {\bf B550} (1999) 151--168,
  [\href{http://arxiv.org/abs/hep-th/9812219}{{\tt hep-th/9812219}}].

\bibitem{Schomerus:1999ug}
V.~Schomerus, {\it {D-branes and deformation quantization}},  {\em JHEP} {\bf
  9906} (1999) 030, [\href{http://arxiv.org/abs/hep-th/9903205}{{\tt
  hep-th/9903205}}].

\bibitem{Seiberg:2000ms}
N.~Seiberg, L.~Susskind, and N.~Toumbas, {\it {Strings in background electric
  field, space / time noncommutativity and a new noncritical string theory}},
  {\em JHEP} {\bf 0006} (2000) 021,
  [\href{http://arxiv.org/abs/hep-th/0005040}{{\tt hep-th/0005040}}].

\bibitem{Gopakumar:2000na}
R.~Gopakumar, J.~M. Maldacena, S.~Minwalla, and A.~Strominger, {\it {S duality
  and noncommutative gauge theory}},  {\em JHEP} {\bf 0006} (2000) 036,
  [\href{http://arxiv.org/abs/hep-th/0005048}{{\tt hep-th/0005048}}].

\bibitem{Cornalba:2001sm}
L.~Cornalba and R.~Schiappa, {\it {Nonassociative star product deformations for
  D-brane world volumes in curved backgrounds}},  {\em Commun.Math.Phys.} {\bf
  225} (2002) 33--66, [\href{http://arxiv.org/abs/hep-th/0101219}{{\tt
  hep-th/0101219}}].

\bibitem{Chu:2002in}
C.-S. Chu and P.-M. Ho, {\it {Noncommutative D-brane and open string in pp wave
  background with B field}},  {\em Nucl.Phys.} {\bf B636} (2002) 141--158,
  [\href{http://arxiv.org/abs/hep-th/0203186}{{\tt hep-th/0203186}}].

\bibitem{Alekseev:1999bs}
A.~Y. Alekseev, A.~Recknagel, and V.~Schomerus, {\it {Noncommutative world
  volume geometries: Branes on SU(2) and fuzzy spheres}},  {\em JHEP} {\bf
  9909} (1999) 023, [\href{http://arxiv.org/abs/hep-th/9908040}{{\tt
  hep-th/9908040}}].

\bibitem{Alekseev:2000fd}
A.~Y. Alekseev, A.~Recknagel, and V.~Schomerus, {\it {Brane dynamics in
  background fluxes and noncommutative geometry}},  {\em JHEP} {\bf 0005}
  (2000) 010, [\href{http://arxiv.org/abs/hep-th/0003187}{{\tt
  hep-th/0003187}}].

\bibitem{Ho:2000fv}
P.-M. Ho and Y.-T. Yeh, {\it {Noncommutative D-brane in nonconstant NS-NS B
  field background}},  {\em Phys.Rev.Lett.} {\bf 85} (2000) 5523--5526,
  [\href{http://arxiv.org/abs/hep-th/0005159}{{\tt hep-th/0005159}}].

\bibitem{Majid:1994cy}
S.~Majid and H.~Ruegg, {\it {Bicrossproduct structure of kappa Poincare group
  and noncommutative geometry}},  {\em Phys.Lett.} {\bf B334} (1994) 348--354,
  [\href{http://arxiv.org/abs/hep-th/9405107}{{\tt hep-th/9405107}}].

\bibitem{Lukierski:1991pn}
J.~Lukierski, H.~Ruegg, A.~Nowicki, and V.~N. Tolstoi, {\it {Q deformation of
  Poincare algebra}},  {\em Phys.Lett.} {\bf B264} (1991) 331--338.

\bibitem{Lukierski:1992dt}
J.~Lukierski, A.~Nowicki, and H.~Ruegg, {\it {New quantum Poincare algebra and
  k deformed field theory}},  {\em Phys.Lett.} {\bf B293} (1992) 344--352.

\bibitem{AmelinoCamelia:2000mn}
G.~Amelino-Camelia, {\it {Relativity in space-times with short distance
  structure governed by an observer independent (Planckian) length scale}},
  {\em Int.J.Mod.Phys.} {\bf D11} (2002) 35--60,
  [\href{http://arxiv.org/abs/gr-qc/0012051}{{\tt gr-qc/0012051}}].

\bibitem{AmelinoCamelia:2000ge}
G.~Amelino-Camelia, {\it {Testable scenario for relativity with minimum
  length}},  {\em Phys.Lett.} {\bf B510} (2001) 255--263,
  [\href{http://arxiv.org/abs/hep-th/0012238}{{\tt hep-th/0012238}}].

\bibitem{KowalskiGlikman:2004qa}
J.~Kowalski-Glikman, {\it {Introduction to doubly special relativity}},  {\em
  Lect.Notes Phys.} {\bf 669} (2005) 131--159,
  [\href{http://arxiv.org/abs/hep-th/0405273}{{\tt hep-th/0405273}}].

\bibitem{AmelinoCamelia:2001fd}
G.~Amelino-Camelia and M.~Arzano, {\it {Coproduct and star product in field
  theories on Lie algebra noncommutative space-times}},  {\em Phys.Rev.} {\bf
  D65} (2002) 084044, [\href{http://arxiv.org/abs/hep-th/0105120}{{\tt
  hep-th/0105120}}].

\bibitem{Dimitrijevic:2003wv}
M.~Dimitrijevic, L.~Jonke, L.~Moller, E.~Tsouchnika, J.~Wess, et~al., {\it
  {Deformed field theory on kappa space-time}},  {\em Eur.Phys.J.} {\bf C31}
  (2003) 129--138, [\href{http://arxiv.org/abs/hep-th/0307149}{{\tt
  hep-th/0307149}}].

\bibitem{Agostini:2004cu}
A.~Agostini, G.~Amelino-Camelia, M.~Arzano, and F.~D'Andrea, {\it {Action
  functional for kappa-Minkowski noncommutative spacetime}},
  \href{http://arxiv.org/abs/hep-th/0407227}{{\tt hep-th/0407227}}.

\bibitem{Schenkel:2010sc}
A.~Schenkel and C.~F. Uhlemann, {\it {Field Theory on Curved Noncommutative
  Spacetimes}},  {\em SIGMA} {\bf 6} (2010) 061,
  [\href{http://arxiv.org/abs/1003.3190}{{\tt arXiv:1003.3190}}].

\bibitem{Meljanac:2011cs}
S.~Meljanac, A.~Samsarov, J.~Trampetic, and M.~Wohlgenannt, {\it {Scalar field
  propagation in the $\phi^4$ kappa-Minkowski model}},  {\em JHEP} {\bf 1112}
  (2011) 010, [\href{http://arxiv.org/abs/1111.5553}{{\tt arXiv:1111.5553}}].

\bibitem{Dimitrijevic:2003pn}
M.~Dimitrijevic, F.~Meyer, L.~Moller, and J.~Wess, {\it {Gauge theories on the
  kappa Minkowski space-time}},  {\em Eur.Phys.J.} {\bf C36} (2004) 117--126,
  [\href{http://arxiv.org/abs/hep-th/0310116}{{\tt hep-th/0310116}}].

\bibitem{Dimitrijevic:2005xw}
M.~Dimitrijevic, L.~Jonke, and L.~Moller, {\it {U(1) gauge field theory on
  kappa-Minkowski space}},  {\em JHEP} {\bf 0509} (2005) 068,
  [\href{http://arxiv.org/abs/hep-th/0504129}{{\tt hep-th/0504129}}].

\bibitem{Borowiec:2013lca}
A.~Borowiec and A.~Pachol, {\it {Unified description for $\kappa$-deformations
  of orthogonal groups}},  {\em Eur. Phys. J.} {\bf C74} (2014), no.~3 2812,
  [\href{http://arxiv.org/abs/1311.4499}{{\tt arXiv:1311.4499}}].

\bibitem{Juric:2015hda}
T.~Juri\'c, S.~Meljanac, and A.~Samsarov, {\it {Light-like
  $\kappa$-deformations and scalar field theory via Drinfeld twist}},
  \href{http://arxiv.org/abs/1506.02475}{{\tt arXiv:1506.02475}}.

\bibitem{Dimitrijevic:2011jg}
M.~Dimitrijevic and L.~Jonke, {\it {A Twisted look on kappa-Minkowski: U(1)
  gauge theory}},  {\em JHEP} {\bf 1112} (2011) 080,
  [\href{http://arxiv.org/abs/1107.3475}{{\tt arXiv:1107.3475}}].

\bibitem{Borowiec:2013gca}
A.~Borowiec, J.~Lukierski, and A.~Pacho\l{}, {\it {Twisting and
  $\kappa$-Poincar\'e}},  {\em J.Phys.} {\bf A47} (2014), no.~40 405203,
  [\href{http://arxiv.org/abs/1312.7807}{{\tt arXiv:1312.7807}}].

\bibitem{Borowiec:2008uj}
A.~Borowiec and A.~Pacho\l{}, {\it {kappa-Minkowski spacetime as the result of
  Jordanian twist deformation}},  {\em Phys.Rev.} {\bf D79} (2009) 045012,
  [\href{http://arxiv.org/abs/0812.0576}{{\tt arXiv:0812.0576}}].

\bibitem{Matsumoto:2015ypa}
T.~Matsumoto, D.~Orlando, S.~Reffert, J.-i. Sakamoto, and K.~Yoshida, {\it
  {Yang-Baxter deformations of Minkowski spacetime}},
  \href{http://arxiv.org/abs/1505.04553}{{\tt arXiv:1505.04553}}.

\bibitem{Metsaev:1998it}
R.~Metsaev and A.~A. Tseytlin, {\it {Type IIB superstring action in AdS(5) x
  S**5 background}},  {\em Nucl.Phys.} {\bf B533} (1998) 109--126,
  [\href{http://arxiv.org/abs/hep-th/9805028}{{\tt hep-th/9805028}}].

\bibitem{Bena:2003wd}
I.~Bena, J.~Polchinski, and R.~Roiban, {\it Hidden symmetries of the
  $\mathit{AdS}_{5}\times \mathit{S}^5$ superstring},  {\em Phys. Rev.} {\bf
  D69} (2004) 046002, [\href{http://arxiv.org/abs/hep-th/0305116}{{\tt
  hep-th/0305116}}].

\bibitem{Hashimoto:1999ut}
A.~Hashimoto and N.~Itzhaki, {\it {Noncommutative Yang-Mills and the AdS / CFT
  correspondence}},  {\em Phys.Lett.} {\bf B465} (1999) 142--147,
  [\href{http://arxiv.org/abs/hep-th/9907166}{{\tt hep-th/9907166}}].

\bibitem{Maldacena:1999mh}
J.~M. Maldacena and J.~G. Russo, {\it {Large N limit of noncommutative gauge
  theories}},  {\em JHEP} {\bf 9909} (1999) 025,
  [\href{http://arxiv.org/abs/hep-th/9908134}{{\tt hep-th/9908134}}].

\bibitem{Alday:2005ww}
L.~F. Alday, G.~Arutyunov, and S.~Frolov, {\it {Green-Schwarz strings in
  TsT-transformed backgrounds}},  {\em JHEP} {\bf 0606} (2006) 018,
  [\href{http://arxiv.org/abs/hep-th/0512253}{{\tt hep-th/0512253}}].

\bibitem{Matsumoto:2014cja}
T.~Matsumoto and K.~Yoshida, {\it {Integrable deformations of the AdS$_{5}
  \times S^5$ superstring and the classical Yang-Baxter equation $- Towards$
  $the$ $gravity/CYBE$ $correspondence -$}},  {\em J.Phys.Conf.Ser.} {\bf 563}
  (2014), no.~1 012020, [\href{http://arxiv.org/abs/1410.0575}{{\tt
  arXiv:1410.0575}}].

\bibitem{Chari}
V.~Chari and A.~Pressley, {\it {A Guide To Quantum Groups}},  {\em Cambridge,
  UK: Univ. Press} (1994).

\bibitem{Reshetikhin:1990ep}
N.~Reshetikhin, {\it {Multiparameter quantum groups and twisted quasitriangular
  Hopf algebras}},  {\em Lett.Math.Phys.} {\bf 20} (1990) 331--335.

\bibitem{Giaquinto:1994jx}
A.~Giaquinto and J.~J. Zhang, {\it {Bialgebra actions, twists, and universal
  deformation formulas}},  {\em J.Pure Appl.Algebra} {\bf 128} (1998) 133--151,
  [\href{http://arxiv.org/abs/hep-th/9411140}{{\tt hep-th/9411140}}].

\bibitem{Tolstoy:2008zz}
V.~Tolstoy, {\it {Twisted quantum deformations of Lorentz and Poincare
  algebras}},  {\em Bulg.J.Phys.} {\bf 35} (2008) 441--459,
  [\href{http://arxiv.org/abs/0712.3962}{{\tt arXiv:0712.3962}}].

\bibitem{Kulish:1998be}
P.~Kulish, V.~Lyakhovsky, and A.~Mudrov, {\it {Extended jordanian twists for
  Lie algebras}},  {\em J.Math.Phys.} {\bf 40} (1999) 4569,
  [\href{http://arxiv.org/abs/math/9806014}{{\tt math/9806014}}].

\bibitem{Beisert:2005if}
N.~Beisert and R.~Roiban, {\it {Beauty and the twist: The Bethe ansatz for
  twisted N=4 SYM}},  {\em JHEP} {\bf 0508} (2005) 039,
  [\href{http://arxiv.org/abs/hep-th/0505187}{{\tt hep-th/0505187}}].

\bibitem{vanTongeren:2013gva}
S.~J. van Tongeren, {\it {Integrability of the $\ads$ superstring and its
  deformations}},  {\em J.Phys.} {\bf A47} (2014), no.~43 433001,
  [\href{http://arxiv.org/abs/1310.4854}{{\tt arXiv:1310.4854}}].

\bibitem{Kawaguchi:2013lba}
I.~Kawaguchi, T.~Matsumoto, and K.~Yoshida, {\it {Schroedinger sigma models and
  Jordanian twists}},  {\em JHEP} {\bf 1308} (2013) 013,
  [\href{http://arxiv.org/abs/1305.6556}{{\tt arXiv:1305.6556}}].

\bibitem{Matsumoto:2015jja}
T.~Matsumoto and K.~Yoshida, {\it {Yang-Baxter sigma models based on the
  CYBE}},  {\em Nucl.Phys.} {\bf B893} (2015) 287--304,
  [\href{http://arxiv.org/abs/1501.03665}{{\tt arXiv:1501.03665}}].

\bibitem{Leigh:1995ep}
R.~G. Leigh and M.~J. Strassler, {\it {Exactly marginal operators and duality
  in four-dimensional N=1 supersymmetric gauge theory}},  {\em Nucl.Phys.} {\bf
  B447} (1995) 95--136, [\href{http://arxiv.org/abs/hep-th/9503121}{{\tt
  hep-th/9503121}}].

\bibitem{Matsumoto:2015uja}
T.~Matsumoto and K.~Yoshida, {\it {Schr\"odinger geometries arising from
  Yang-Baxter deformations}},  \href{http://arxiv.org/abs/1502.00740}{{\tt
  arXiv:1502.00740}}.

\bibitem{Dasgupta:2001zu}
K.~Dasgupta and M.~Sheikh-Jabbari, {\it {Noncommutative dipole field
  theories}},  {\em JHEP} {\bf 0202} (2002) 002,
  [\href{http://arxiv.org/abs/hep-th/0112064}{{\tt hep-th/0112064}}].

\bibitem{Hashimoto:2005hy}
A.~Hashimoto and K.~Thomas, {\it {Non-commutative gauge theory on d-branes in
  melvin universes}},  {\em JHEP} {\bf 0601} (2006) 083,
  [\href{http://arxiv.org/abs/hep-th/0511197}{{\tt hep-th/0511197}}].

\bibitem{Doplicher:1994tu}
S.~Doplicher, K.~Fredenhagen, and J.~E. Roberts, {\it {The Quantum structure of
  space-time at the Planck scale and quantum fields}},  {\em Commun.Math.Phys.}
  {\bf 172} (1995) 187--220, [\href{http://arxiv.org/abs/hep-th/0303037}{{\tt
  hep-th/0303037}}].

\bibitem{Bahns:2002vm}
D.~Bahns, S.~Doplicher, K.~Fredenhagen, and G.~Piacitelli, {\it {On the
  Unitarity problem in space-time noncommutative theories}},  {\em Phys.Lett.}
  {\bf B533} (2002) 178--181, [\href{http://arxiv.org/abs/hep-th/0201222}{{\tt
  hep-th/0201222}}].

\bibitem{Aharony:2000gz}
O.~Aharony, J.~Gomis, and T.~Mehen, {\it {On theories with lightlike
  noncommutativity}},  {\em JHEP} {\bf 0009} (2000) 023,
  [\href{http://arxiv.org/abs/hep-th/0006236}{{\tt hep-th/0006236}}].

\bibitem{Hashimoto:2002nr}
A.~Hashimoto and S.~Sethi, {\it {Holography and string dynamics in time
  dependent backgrounds}},  {\em Phys.Rev.Lett.} {\bf 89} (2002) 261601,
  [\href{http://arxiv.org/abs/hep-th/0208126}{{\tt hep-th/0208126}}].

\bibitem{Lukierski:2002ii}
J.~Lukierski, V.~Lyakhovsky, and M.~Mozrzymas, {\it {Kappa deformations of D =
  4 Weyl and conformal symmetries}},  {\em Phys.Lett.} {\bf B538} (2002)
  375--384, [\href{http://arxiv.org/abs/hep-th/0203182}{{\tt hep-th/0203182}}].

\bibitem{zakrzewski:1994}
S.~Zakrzewski, {\it Poisson structures on the lorentz group},  {\em Letters in
  Mathematical Physics} {\bf 32} (1994), no.~1 11--23.

\bibitem{Frolov:2005iq}
S.~Frolov, R.~Roiban, and A.~A. Tseytlin, {\it {Gauge-string duality for
  (non)supersymmetric deformations of N=4 super Yang-Mills theory}},  {\em
  Nucl.Phys.} {\bf B731} (2005) 1--44,
  [\href{http://arxiv.org/abs/hep-th/0507021}{{\tt hep-th/0507021}}].

\bibitem{Fokken:2013aea}
J.~Fokken, C.~Sieg, and M.~Wilhelm, {\it {Non-conformality of ${{\gamma
  }_{i}}$-deformed N = 4 SYM theory}},  {\em J.Phys.} {\bf A47} (2014) 455401,
  [\href{http://arxiv.org/abs/1308.4420}{{\tt arXiv:1308.4420}}].

\bibitem{Fokken:2014soa}
J.~Fokken, C.~Sieg, and M.~Wilhelm, {\it {A piece of cake: the ground-state
  energies in $\gamma_{i}$ -deformed $ \mathcal{N} $ = 4 SYM theory at leading
  wrapping order}},  {\em JHEP} {\bf 1409} (2014) 78,
  [\href{http://arxiv.org/abs/1405.6712}{{\tt arXiv:1405.6712}}].

\bibitem{Borowiec:2008se}
A.~Borowiec, J.~Lukierski, and V.~Tolstoy, {\it {New twisted quantum
  deformations of D=4 super-Poincare algebra}},
  \href{http://arxiv.org/abs/0803.4167}{{\tt arXiv:0803.4167}}.

\bibitem{Berkovits:2008ic}
N.~Berkovits and J.~Maldacena, {\it {Fermionic T-Duality, Dual Superconformal
  Symmetry, and the Amplitude/Wilson Loop Connection}},  {\em JHEP} {\bf 0809}
  (2008) 062, [\href{http://arxiv.org/abs/0807.3196}{{\tt arXiv:0807.3196}}].

\bibitem{Beisert:2008iq}
N.~Beisert, R.~Ricci, A.~A. Tseytlin, and M.~Wolf, {\it {Dual Superconformal
  Symmetry from AdS(5) x S**5 Superstring Integrability}},  {\em Phys.Rev.}
  {\bf D78} (2008) 126004, [\href{http://arxiv.org/abs/0807.3228}{{\tt
  arXiv:0807.3228}}].

\bibitem{zakrzewski:1997}
S.~Zakrzewski, {\it Poisson structures on the poincar\'{e} group},  {\em
  Communications in Mathematical Physics} {\bf 185} (1997), no.~2 285--311,
  [\href{http://arxiv.org/abs/q-alg/9602001}{{\tt q-alg/9602001}}].

\end{thebibliography}\endgroup

\end{document}